\documentclass[11pt]{JHEP3}

\usepackage{epsfig,multicol,bbm}
\usepackage{cite}
\usepackage{amsmath,amsfonts,amssymb}
\usepackage{array}
\usepackage{graphicx}
\usepackage{subfig}
\usepackage{enumerate}
\usepackage{bm}
\usepackage[utf8]{inputenc}
\usepackage{cancel}
\usepackage[british]{babel}

\DeclareMathAlphabet{\mathpzc}{OT1}{pzc}{m}{it}

\input epsf.sty

\def\be{\begin{equation}} 
\def\ee{\end{equation}}
\def\ba{\begin{array}}
\def\bacc{\begin{array} {cc}}
\def\ea{\end{array}}
\def\bea{\begin{eqnarray}}
\def\eea{\end{eqnarray}}
\def\bd{\begin{displaymath}}
\def\ed{\end{displaymath}}
\def\calN{{\mathcal N}}
\def\calF{{\mathcal F}}
\def\calS{{\mathcal S}}

\def\calL{{\mathcal L}}
\def\S{{\mathcal S}}

\def\V{{\mathcal V}}

\def\bD{{\overline{\textrm{D}}}}

\def\bDt{{\overline{\textrm{D}3}}}
\def\bG{{\bar{G}}}
\def\bS{{\bar{S}}}
\def\bX{{\bar{X}}}

\def\blambda{{\bar{\lambda}}}

\def\bpsi{{\bar{\psi}}}

\def\btau{{\bar{\tau}}}

\def\bT{{\bar{T}}}
\def\bS{{\bar{S}}}

\def\b1{{\bar{1}}}
\def\Z{\mathbb Z}
\def\R{\mathbb R}

\def\Id{\mathbbm 1}

\def\i{{\rm i}}

\def\nn{\nonumber}

\def\Mpl{M_{Pl}}

\def\pd{\partial}
\def\Im{{\rm{Im}}}
\def\Re{{\rm{Re}}}

\def\d{{\rm{d}}}

\def\dofs{{\it d.o.f. }}

\usepackage{cancel, color}

\title{Anti-D3 branes and moduli in non-linear supergravity}

\author{Maria P. Garcia del Moral$^a$, Susha Parameswaran$^{b}$, Norma Quiroz$^c$, Ivonne Zavala$^d$  \\ \\
${}^a$Departamento de Fisica, Universidad de Antofagasta, Aptdo 02800, Chile\\
${}^b$Department of Mathematical Sciences, University of Liverpool, Liverpool, L69 7ZL, UK\\
${}^c${Departamento de Ciencias de la Naturaleza,
CUSUR, Universidad de Guadalajara\\
Enrique Arreola Silva 883, C.P. 49000, Cd.Guzm\'an, Jalisco, M\'exico}\\
${}^d$Department of Physics, Swansea University, Swansea, SA2 8PP, UK\\

\vspace{.5cm}
E-mail: {\email{maria.garciadelmoral@uantof.cl}\,,
\email{Susha.Parameswaran@liverpool.ac.uk}\,,
\email{norma.quiroz@cusur.udg.mx}\,,
\email{e.i.zavalacarrasco@swansea.ac.uk}}
}

\abstract{Anti-D3 branes and non-perturbative effects in flux compactifications spontaneously break supersymmetry and stabilise moduli in a metastable de Sitter vacua.  The low energy 4D effective field theory description for such models would be a supergravity theory with non-linearly realised supersymmetry.   Guided by string theory modular symmetry, we compute this non-linear supergravity theory, including dependence on all bulk moduli.  Using either a constrained chiral superfield or a constrained vector field, the uplifting contribution to the scalar potential from the anti-D3 brane can be parameterised either as an F-term or Fayet-Iliopoulos D-term.
Using again the modular symmetry, we show that 4D non-linear supergravities that descend from string theory have an enhanced protection from quantum corrections by non-renormalisation theorems.  The superpotential giving rise to metastable de Sitter vacua is robust against perturbative string-loop and $\alpha'$  corrections.

\vspace{1cm}

 }


\begin{document}

\section{Introduction}
One of the main challenges in connecting String Theory to our observed Universe is to provide a string theoretic description of the early and late time accelerated expansions.  This requires us to identify well-controlled string theory vacua whose 4D geometry corresponds to de Sitter (dS) or quasi-dS, with all moduli stabilised.  Moduli stabilisation into a dS vacuum has been notoriously difficult to achieve, and the no-go theorems \cite{Gibbons, MN} made clear what ingredients would be necessary.  In particular, taking the classical two-derivative 10D string supergravities, including localised D$p$-brane, $\overline{{\rm D}p}$-branes and O$p$-plane sources, the Einstein's and dilaton equations imply that one needs negative tensions and negative internal curvature to source dS.  A way to evade these restrictions is to include higher derivative corrections.

Although this makes the explicit construction of dS vacua -- and moreover metastable dS vacua -- difficult, several mechanisms have been proposed.  Arguably the most used construction is to uplift\footnote{Alternative methods include D-term uplifting via gauge fluxes on wrapped D7-branes \cite{Burgess:2003ic, Cremades:2007ig}, F-term uplifting via complex structure \cite{Saltman:2004sn}, $\alpha'$ corrections to the K\"ahler potential in no-scale flux compactifications \cite{Balasubramanian:2004uy} and F-term uplifting from dilaton dependent non-perturbative terms \cite{Cicoli:2012fh}.} a Minkowski or adS vacuum to dS with the addition of a positive energy density from an $\bDt$-brane.  For a small number of probe $\bDt$-branes at the tip of a highly warped throat, an effective field theory analysis shows that such a configuration is metastable \cite{Kachru:2002gs}.   There is a non-perturbative instability to antibrane-flux annihilation, but the timescale of this stability can be far longer than the age of the Universe.  Moreover, if we place the $\bDt$-brane on top of an O3-plane, then any concerns about tachyonic instabilities that might appear when going beyond the probe approximation (see \cite{Polchinski:2015bea, Bena:2016fqp} and references therein) are simply projected out.

The original $\bDt$-brane uplift scenario, by Kachru, Kallosh, Linde and Trivedi (KKLT) \cite{KKLT}, was presented in three steps.  Firstly, a Giddings, Kachru, Polchinski (GKP) \cite{GKP} type IIB flux compactification stabilises the dilaton and complex structure moduli in a non-supersymmetric vacuum.  Next, the resulting runaway in the K\"ahler modulus is stabilised into a supersymmetry restoring vacuum by non-perturbative effects, such as gaugino condensation on wrapped D7-branes and/or Euclidean D3-branes.  Finally, the supersymmetric adS vacuum is uplifted to a supersymmetry breaking dS vacuum by the $\bDt$-brane.  Note that the dS vacuum is achieved with a combination of the $\bDt$-brane and non-perturbative effects -- without the non-perturbative effects, the anti-brane would just give a runaway towards decompactification -- so, as expected, quantum corrections are essential to evade the dS no-go theorems. 

The $\bDt$-brane, as well as uplifting the classical vacuum energy to dS, spontaneously breaks supersymmetry.  Any string compactification with spontaneously broken supersymmetry would have a non-linearly realised local supersymmetry (``non-linear supergravity'') as its effective field theory description at energies below\footnote{At energies above the visible sector superpartner masses, the latter can still be parameterised by soft susy breaking terms.} the mass of the goldstino's superpartner (usually the sgoldstino).  That is, the action is invariant under non-linear supersymmetry transformations, and the non-linear supersymmetry transformation for the goldstino implies that all solutions spontaneously break supersymmetry.  The goldstino is eaten by the gravitino in the super-Higgs mechanism.  Non-linear supergravity can be written in terms of non-linear or constrained supermultiplets, which contain a single elementary field (either bosonic or fermionic) and the goldstino.  This superfield description makes it easy to couple to supergravity and matter, starting with \cite{UlfRocek, SW}.  Recently, \cite{PuredS, Kallosh:2015tea, Kallosh:2016dcq, Bergshoeff:2016psz} computed the component form for supergravity coupled to a nilpotent chiral superfield, $S^2=0$, which carries the goldstino, and general matter.

Although the original KKLT construction parameterised the $\bDt$-brane contribution to the 4D low energy effective field theory (LEEFT) in terms of an uplift term that explicitly broke supersymmetry, the connection between non-linear supergravity and $\bDt$-branes has been explored a lot recently in an effort to find a well-controlled LEEFT description of KKLT setup.   Notably, by studying an $\bDt$-brane placed on top of an O3-plane in a warped throat geometry, the massless degrees of freedom on the brane were identified with the goldstino of spontaneously broken supersymmetry, which can be described by a nilpotent chiral superfield $S^2=0$ \cite{McGuirk:2012sb, KW, BDKVW, KQU, Garcia-Etxebarria:2015lif}.  Motivated by matching the degrees of freedom, canonical kinetic terms and scalar potential arising from the $\bDt$-brane action in a GKP background, a non-linear supergravity theory was proposed for the final step of the KKLT scenario \cite{BDKVW}.  In detail, for a KKLT variant with supersymmetric background fluxes and the volume modulus stabilised by racetrack non-perturbative effects:
\bea
&&K = -3\ln(T+\bT - S \bS) \quad \textrm{and} \quad W = A e^{-a T} + B e^{-b T} + M^2 S \quad \textrm{and} \quad S^2 = 0 \,, 
\eea
which gives:
\bea
&& V = \frac{M^4-3|W_0|^2}{(T_0+\bar{T_0})^2} \quad \textrm{with} \quad W_0=W(T_0) 
\eea
and $T_0$ the value of $T$ at its minimum, where $D_T W|_{T=T_0}=0$. 

The purpose of this paper is two-fold.  First, we provide the full 4D $\calN=1$ non-linear supergravity action which describes at low energies an $\bDt$-brane in a GKP flux compactification including non-perturbative effects {\it a.k.a.}~the KKTL scenario.  In particular, we go beyond the original three-step process and include the $\bDt$-brane from the beginning, and thus its couplings to all the bulk moduli.  Indeed, intuitively, one imagines the $\bDt$-brane to have emerged, somehow, together with the 10D compactification and background fluxes.  This picture allows us to order the dynamics in terms of energy scales, with the mass hierarchy set up to be:
\be
 M_{pl} \gtrsim M_{s} \gg M^{w}_{s} \sim M_{\bcancel{susy}}  \gg  M^{w}_{kk} \sim  \Lambda \sim \Lambda_{np} \gg M_{\tau ,Z} \gg  M_{T}  \sim M_{3/2} \sim M_{goldstino} 
\ee
where the tension of the $\bDt$-brane goes as the warped string scale, $M^{w}_{s}$, $\Lambda_{np}$ is the scale at which non-perturbative effects kick in, and $\Lambda$ is the UV cutoff for the 4D LEEFT.  As we explain below, the supersymmetry breaking scale, $M_{\bcancel{susy}}$, is associated with the tension of the $\bDt$-brane, which - being placed in an orientifold flux compactification - only realises supersymmetry non-linearly.  The 4D goldstino descends from the worldvolume fermion on the $\bDt$-brane, and -- due to the non-linear realisation of supersymmetry -- has no superpartner.  In the decompactification limit, supersymmetry is still broken due to the interplay between the $\bDt$-brane and the orientifold.  Supersymmetry would only be restored via the non-perturbative process of antibrane-flux annihilation, where the supersymmetry breaking degrees of freedom on the $\bDt$-brane are replaced by the supersymmetric degrees of freedom of a stack of D3 branes.

We are able to write down the full 4D LEEFT by using a combination of dimensional reduction, non-linear supersymmetry and remnants of the modular invariance of 10D type IIB string theory.  The non-linear supergravity action can be written in terms of a real K\"ahler potential, holomorphic superpotential and gauge kinetic functions, and Fayet-Iliopoulos terms.  One interesting observation is that there are several equivalent ways to express the action.  For example, we can place the goldstino in a nilpotent chiral superfield \cite{Komargodski:2009rz}, $S^2=0$, a constrained chiral superfield \cite{UlfRocek}, $X$, which obeys the nilpotency plus a derivative constraint, or a constrained vector superfield \cite{SW}, $V$.  The uplift term associated to the $\bDt$-brane therefore corresponds either to an F-term or an FI D-term, even though the D-term associated to the constrained vector superfield is not associated with any gauge symmetry.

The second objective of the paper is to study the robustness of the uplifted 4D dS vacuum against quantum gravity corrections in $g_s$ and $\alpha'$.  This is important, especially because the dS vacuum is obtained by invoking classical and non-perturbative effects, so one has to check that the vacuum is not destabilised by perturbative corrections, which would dominate over non-perturbative effects.  Early arguments on using non-perturbative effects to stabilise flat directions in CY compactifications relied on non-renormalisation theorems for the holomorphic superpotential \cite{Witten:1985bz, Dine:1986vd}.  Certain Peccei-Quinn (PQ) shift symmetries forbade the superpotential from depending on the axionic partners of the dilaton and volume modulus, and then holomorphicity of the superpotential forbade corrections to all finite orders in $g_s$ and $\alpha'$.  In flux compactifications non-renormalisation is not so simple, as the superpotential does depend explicitly on the dilaton through the Gukov, Vafa, Witten (GVW) superpotential \cite{Gukov:1999ya}. However, the non-renormalisation theorem was extended to this case in \cite{BEQ}.  We here extend further the non-renormalisation theorem to compactifications in the presence of an $\bDt$-brane, by using -- as in \cite{BEQ} - remnants of the stringy modular symmetry spontaneously broken by fluxes and the nilpotential superfield.
Although the K\"ahler potential will receive order-by-order perturbative corrections in $g_s$ and $\alpha'$, as the vacuum structure is generally determined by the superpotential (see however, \cite{Balasubramanian:2004uy, Balasubramanian:2005zx}), non-renormalisation of the latter offers an enhanced robustness to the ``uplifted'' dS vacua against quantum gravity corrections. 

The paper is organised as follows.  In Section 2, we review how $\bDt$-branes in Calabi-Yau orientifold flux compactifications spontaneously break supersymmetry, together with the degrees of freedom and the action of a $\bDt$-brane in such a background.  In Section 3, after reviewing the background material on constrained superfields, we derive the 4D LEEFT describing the KKLT setup, parameterising the $\bDt$-brane uplift both as an F-term and a D-term potential.  Section 4 gives a proof that the superpotential found does not receive any corrections to all finite orders in $g_s$ and $\alpha'$.  We summarise our results and some open questions in Section 5.  Two appendices give our conventions and a derivation of the modular transformation properties of the worldvolume fermion on the $\bDt$-brane.  An expert reader can simply jump to our main results from Section \ref{S:NLsugraKKLT} onwards.

\section{Spontaneous supersymmetry breaking by $\bDt$-branes}

The purpose of this section is to review how $\bDt$-branes spontaneously break supersymmetry in a Calabi-Yau orientifold flux compactification.  After introducing the type IIB supergravity theory, including localised sources, we first show how an $\bDt$-brane spontaneously breaks the supersymmetry of a flat orientifold background.  Correspondingly, its worldvolume action reduces to a Volkov-Akulov theory, with non-linearly realised global supersymmetry.  We then build on these results, to review how an $\bDt$-brane spontaneously breaks the supersymmetry of a Calabi-Yau flux background, with supersymmetry preserving background fluxes.  We present the leading order worldvolume action in this background, which was worked out in \cite{BDKVW}.  In the next section we will work out how this action contributes to the non-linear supergravity theory corresponding to a Calabi-Yau flux compactification with a probe $\bDt$-brane.

\subsection{Setup}

\label{S:setup}

Our starting point is type IIB supergravity, which has 16 + 16 linearly realised supersymmetries.  The degrees of freedom are the graviton $g_{MN}$, axio-dilaton $\tau =  C_0 + \i \, e^{-\phi}$, three-form RR flux $F_{(3)}= \d C_2 $, three-form NS-NS flux $H_{(3)} = \d B_{(2)}$, self-dual five-form RR flux $F_{(5)} =  \d C_4 + \dots$, and their fermionic superpartners; the complex-Weyl gravitino $\Psi_M$ ($M=0, \dots, 9$ and $\Gamma^{11} \Psi_M = -\Psi_M$) and dilatino $\lambda$ ($\Gamma^{11} \lambda = \lambda$).  The action, in the Einstein frame, is:
\bea
S_{IIB} &=& \frac{1}{2\kappa_{10}^2} \int d^{10}x \sqrt{-g} \left[ R - \frac{\pd_M \bar\tau \pd^M\tau}{2(\Im\tau)^2} - \frac{G_{(3)} . \bG_{(3)}}{12\Im\tau} - \frac{1}{480}F_{(5)}^2 \right] \nn \\
&& -\frac{\i}{8\kappa_{10}^2} \int \frac{C_{(4)} \wedge G_{(3)} \wedge \bG_{(3)}}{\Im\tau} + \textrm{fermions} + \textrm{higher derivative corrections} \label{E:IIBaction}
\eea
where $G_{(3)} = F_{(3)} - \tau H_{(3)} $ and $2 \kappa_{10}^2 = (2\pi)^7\alpha'^4$ and ${\alpha'}^{-1/2}=M_s$ is the string scale.

In addition to the bulk 10D supergravity, we consider a number of localised sources; D-branes, $\bD$-branes and O-planes.  These sources each realise half of the bulk spacetime supersymmetries linearly and half non-linearly.   The fields of the 4D worldvolume theory of a D3-brane or $\bDt$-brane are its 10D spacetime coordinates, $X^M$, a $U(1)$ gauge field ${\mathcal A}_\mu$ ($\mu=0,1,2,3$) and a pair  of Majorana-Weyl spinors $\Theta^A$ ($A=1,2$).  For a D-brane, $\Gamma^{11} \Theta^A = -  \Theta^A$, whereas for an $\bD$-brane, $\Gamma^{11} \Theta^A =   \Theta^A$. 

The worldvolume action for D-branes and $\bD$-branes describes the interaction between worldvolume fields and a general supergravity background.  
It can be written as a 4D non-linear sigma model with  curved superspace as the target space, where the worldvolume fields $Z^\Lambda(\sigma) = \left(X^M(\sigma),\Theta^{A}(\sigma) \right)$ ($M= 0,\dots, 9$; $\Lambda = M|A$) define a map from the worldvolume coordinates $\sigma^\mu$ to a superspace with coordinates $Z^\Lambda(\sigma)$  \cite{Cederwall96}.  The action is simply given by the DBI and WZ expressions promoted to superspace; in the Einstein frame we have:
\be
\calS_{3} = -T_3 \int d^4 \sigma \sqrt{- \mbox{det}(\mathpzc{g}_{\mu\nu} + \calF_{\mu\nu})} + q T_3 \int C e^\calF \,, \label{E:S3}
\ee
where the tension of the D3/$\bDt$-brane is given by $T_3=(2\pi)^{-3}\alpha'^{-2}g_s^{-1}$  and $\mathcal F = 2\pi\alpha' \d {\mathcal A} - \mathcal{B}_{(2)}$, with $g_s= \langle e^{\phi}\rangle$ and $\mathcal{B}_{(2)}$  the pull-back to the worldvolume of the spacetime 2-form potential $B_{(2)}$.  Also, for a D3-brane, $q=+1$, whereas an $\bDt$-brane has $q =-1$.  The induced metric on the worldvolume, $\mathpzc{g}_{\mu\nu}$, is expressed in terms of the pullback of the supervielbein $E^{\bar\Lambda}_\Lambda$ (with flat indices $\bar\Lambda$) to the worldvolume:
\be
E_\mu^a(X,\Theta) = \pd_\mu Z^\Lambda E_\Lambda^a(X,\Theta)
\ee
 as:
\be
\mathpzc{g}_{\mu\nu}  \left(X^M(\sigma),\Theta^{A}(\sigma)\right) = E_\mu^a(X,\Theta) E_\nu^b(X,\Theta) \eta_{ab}\,,
\ee
with $\eta_{ab}$ is the 10D Minkowski metric.   Note that the complete, explicit action for D-brane degrees of freedom is only known for a single D-brane or $\bD$-brane in a flat background; for a single brane in a flux background it is known only to quadratic order in the fermions.  However, this will be sufficient for our purposes.

The action is fully covariant under spacetime-local supersymmetry, worldvolume-local general coordinate invariance and a  worldvolume-local fermion symmetry, known as $\kappa$-symmetry, as well as $U(1)$ gauge invariance.  Gauge fixing to the static gauge leaves six scalar degrees of freedom, $X^m$ ($m=4, \dots, 9$), for a D3/$\bDt$-brane, and the spacetime spinors $\Theta^A$ become worldvolume spinors. Gauge-fixing the $\kappa$-symmetry, the local target space supersymmetry combines with a particular $\kappa$-transformation into a global worldvolume supersymmetry, and reduces the 32 fermionic degrees of freedom by half.  Recalling that the equations of motion reduce the fermionic degrees of freedom again to eight, the brane thus has the required supersymmetric matching of the number of fermionic and bosonic (six scalar \dofs and two \dofs from the gauge bosons) physical degrees of freedom. 

In the presence of D3/D7-branes, tadpole cancellation requires the presence of appropriate orientifold O3/O7-planes, whose action is similar to \eqref{E:S3}, with, however, only the rigid pullback of the bulk metric and potential $C_4$ appearing.  The total system is thus defined by:
\be
\calS_{IIB} + \sum_b \calS_{3,7}^b + \sum_o \calS_{3,7}^o
\ee
where $b$ runs over the D3/D7-branes and a single $\bDt$-brane and $o$ runs over the O-planes.

\subsection{$\bDt$-brane in orientifolded flat space}
As a warm up, let us briefly review a probe D3-brane or $\bDt$-brane in a flat orientifolded 4D compactification, taking the global limit \cite{KW}.  The orientifold projects out half of the 16+16 supersymmetries.  The D3-brane  realises the surviving supersymmery linearly and the $\bDt$-brane does so non-linearly.  The D3/$\bDt$-brane scalars are the goldstone bosons of the spontaneously broken translation symmetry and the $\bDt$-brane worldvolume fermions are the goldstinos of spontaneously broken supersymmetry.

In more detail, a convenient choice of orientifolding\footnote{Alternative gauge choices often encountered are setting one of $\Theta^1$ or $\Theta^2$ equal to zero \cite{Schwarz}, or $\Theta^1 = c \Theta^2$ for some constant number $c$ \cite{Grana}.} -- which has to be compatible with the $\kappa$-symmetry gauge fixing \cite{KW} -- is:
\be
 \Theta^1 = \Gamma_{0123} \Theta^2 \,, \quad\textrm{ with } \,\,\Gamma_{0,1,2,3} \,\,\textrm{ flat }
 \,\,\Gamma\textrm{-matrices}
\ee
leaving only one linear combination of the 10D MW spinors, say $\Theta = \Theta^1$, as an independent degree of freedom.    This 10D MW spinor, $\Theta$ (16 \dofs~), can be decomposed into four 4D complex Weyl spinors $\lambda^I$ ($I=0,\dots, 3$)  (4 times 4 real \dofs~).  Indeed, after compactification the structure group reduces as $SO(9,1) \rightarrow SO(3,1)\times SO(6) \rightarrow SO(3,1) \times SU(3)$, under which, for a D-brane:
\be
{\bf 16}' \rightarrow (\bar{\bf{2}}, \bf{4}) \oplus ({\bf{2}}, \bar{\bf{4}}) \rightarrow (\bar{\bf{2}}, \bf{3}) \oplus (\bar{\bf{2}}, \bf{1}) \oplus(\bf{2}, \bar{\bf{3}})  \oplus(\bf{2}, \bar{\bf{1}})\,, \label{E:16bar}
\ee
and for an $\bD$-brane:
\be
{\bf 16} \rightarrow (\bf{2}, \bf{4}) \oplus (\bar{\bf{2}}, \bar{\bf{4}}) \rightarrow ({\bf{2}}, \bf{3}) \oplus ({\bf{2}}, \bf{1}) \oplus(\bar{\bf{2}}, \bar{\bf{3}})  \oplus(\bar{\bf{2}}, \bar{\bf{1}})\,. \label{E:16}
\ee
In both cases, we have a singlet, $\lambda^0$, and a triplet, $\lambda^i$ ($i=1,2,3$), under the $SU(3)$ acting on the complex three-dimensional normal space to the brane. 

If we moreover, place the D3-brane on top of an O3$^-$-plane \cite{Uranga:1999ib}, the worldvolume position scalars, $X^m$ ($m=4,\dots,9$), and the worldvolume gauge bosons, $\mathcal{A}_\mu$, are projected out.  Supersymmetry then implies that their superpartners, the worldvolume spinor, are also projected out.  There are no degrees of freedom left on the D3-brane, and the worldvolume action is vanishing.  An $\bDt$-brane on top of the O3$^-$-plane also carries no worldvolume bosons, meanwhile the worldvolume spinor survives.

Dimensional reduction of the gauged-fixed $\bDt$-brane action \eqref{E:S3} in a flat orientifolded background then directly yields a 4D ${\mathcal N}=4$ Volkov-Akulov action:
\be
\calS_{\bDt} = -M^4 \int d^4\sigma \det E = -M^4 \int E^0 \wedge E^1 \wedge E^2 \wedge E^3 \, , \quad E^\mu = \d \sigma^\mu + \sum_{I=0}^3\bar\lambda^I \gamma^\mu d\lambda^I \,. \label{E:SVAglob}
\ee
where $M^4 = 2T_3$.  The four 4D fermions are the four goldstino fields associated with the 4D ${\mathcal N} = 4$ spontaneously broken supersymmetry. Indeed the action is invariant under a non-linearly realised supersymmetry that acts only on the fermions:
\be
\delta_\epsilon \lambda^I= \epsilon^I  + \sum_{J=0}^3 \left(\blambda^J \gamma^\mu \epsilon^J \right)\pd_\mu \lambda^I\,,
\ee
with $\epsilon^I$ the supersymmetry variation parameter.
 
\subsection{$\bDt$-brane in flux compactifications}

We now consider a probe $\bDt$-brane in an ${\mathcal N}=1$ Calabi-Yau orientifold compactification of type IIB supergravity with background fluxes, known as a GKP compactification \cite{GKP}.  We assume for simplicity that the Calabi-Yau has only one K\"ahler modulus.  
Note also that we choose the background fluxes, complex structure and dilaton to ensure that the flux -- by itself -- would preserve $\calN = 1$ supersymmetry (i.e. fluxes are imaginary self-dual).  This generically induces a potential for the complex structure and dilaton in the 4D effective field theory.

The background geometry is given by the warped metric:
\be
\d s_{10}^2 = e^{2A(y)} \eta_{\mu\nu} \d x^\mu \d x^\nu + e^{-2A(y)} g_{mn} \d y^m \d y^n
\ee
and is supported by a five-form flux:
\be
\tilde F_{(5)} = (1+ *) \left[ \d \alpha(y) \wedge \d x^0 \wedge \d x^1 \wedge \d x^2 \wedge \d x^3 \right] \quad \textrm{with} \quad \alpha(y) = e^{4A(y)}
\ee
and an imaginary self-dual three-form flux threading three-cycles in the internal Calabi-Yau:
\be
*_6 G_{(3)} = \i G_{(3)} \,. 
\ee
The Bianchi identity: 
\be
\d\tilde F_{(5)} = H_{(3)} \wedge F_{(3)} + 2\kappa_{10}^{\,\,2} T_3 \rho_3^{loc}
\ee
is satisfied by including a number of source O-planes together with the fluxes, D-branes and $\overline{\textrm{D}}$-branes, where $\rho_3^{loc}$ is the D3 charge density form from the localized sources.  The orientifold projects out half of the 10D complex Weyl gravitino, $\Psi_M$, similarly to what we have seen for 10D MW worldvolume fermions, $\Theta^A$. The surviving components, with which we may compose a 10D MW spinor\footnote{Two 10D MW spinors can be obtained from a complex Weyl spinor as $\Psi^1 = \frac12 (\Psi + \Psi^C)$ and $\Psi^2 = \frac{1}{2\i} (\Psi - \Psi^C)$ with $\Psi^C \equiv C_{10} \bar \Psi^T$ and $C_{10}$ the 10D charge conjugation matrix.} $\Psi_M$, decomposes into a 4D complex Weyl singlet and 4D complex Weyl triplet with respect to the $SU(3)$ holonomy as in \eqref{E:16bar}.  
  The triplet is massive, leaving one massless complex Weyl gravitino.

In the Calabi-Yau orientifold flux compactification, a probe D-brane (appropriately aligned with the source D-branes) preserves the 4D $\calN=1$ spacetime supersymmetry \cite{FerDav}, whereas an $\bDt$-brane breaks it spontaneously.  For a supersymmetric D-brane (away from the O3$^-$-plane so that worldvolume fields are kept), the singlet $\lambda$ is a 4D gaugino, and joins the worldvolume $U(1)$ gauge field, ${\mathcal A}_\mu$, to form a 4D ${\mathcal N}=1$ vector multiplet.  The three $\psi^i$ each lie in a 4D $\mathcal{N}=1$ chiral supermultiplet, together with a complex scalar field composed from the six worldvolume position fields, $X^m$.  The low energy effective field theory for a D3-brane in a GKP background has been worked out in \cite{Louis}.

Ref. \cite{BDKVW} worked out the worldvolume theory for a probe $\bDt$-brane in the GKP flux background, to lowest order in the fermions.  To simplify the discussion, the brane is again placed on an O3$^-$-plane so that it carries only the worldvolume fermions (their bosonic superpartners being projected out by the orientifold).  The $\bDt$-brane action is obtained from \eqref{E:S3} evaluated in curved superspace, plugging in the GKP background at the position of the brane.  Including the volume fluctuations, $u$, in the metric\footnote{See \cite{Frey:2008xw} for subtleties in identifying the universal K\"ahler modulus in warped compactifications.}, the GKP metric in the 4D Einstein frame is:
\be
ds^2 = e^{2A-6u} g_{\mu\nu} dx^\mu dx^\nu + e^{2u-2A} g_{mn} dy^m dy^n \,. \label{E:metricwv}
\ee
The $\bDt$-brane action in the GKP background, to quadratic order in the spinors and in the Einstein frame, 
is thus computed to be \cite{BDKVW} (compared to Eq. (3.11) in Ref. \cite{BDKVW} we transfer to the Einstein frame\footnote{The relations between 10D string and Einstein frames are $g_{MN\,s} = e^{\phi/2} g_{MN \, E}$, $\Theta_s = e^{\frac{\phi}{8}}\Theta_E$.  Also, in going to the 4D Einstein frame $g_{\mu\nu \, E} = e^{-6u} g_{\mu\nu \, E_4}$, we have $\Theta_E = e^{-3u/2} \Theta_{E_4}$.}, write gamma matrices w.r.t. the unwarped metrics and include the volume fluctuations):
\be
\calL^{\bDt}_{2-f} =\sqrt{-g} \, T_3 e^{4A_0-12u} \bar\Theta \left[ 2 e^{-A_0} \Gamma^\mu \nabla_\mu +\frac{1}{6}e^{\phi/2+3A_0-6u}\left(\Im G^{ISD}_{mnp}\right) \Gamma^{mnp} \right] \Theta \,,\label{E:SbD3quad}
\ee
where the subindex $2-f$ indicates second order in the fermions. The factor $e^{4A_0}$ indicates the warp factor at the position of the $\bDt$-brane, which close to the tip is given by:
\be
e^{4A_0-4u} = r_0^4/L^4
\ee
with $r_0$ the distance between the $\bDt$-brane and the (effective) $N$ D3-branes sourcing the warp factor, measured with the metric $g_{mn}$.  For the $\bDt$-brane in e.g. the deformed conifold, this distance is cut off at  $r_0 \sim Z^{1/3}$, with $Z$ the complex structure modulus associated with the radius of the blown-up $S^3$ in the base of the deformed conifold.  The usual procedure \cite{GKP} is then to exploit the fact that fluxes induce a superpotential which fixes $Z$; indeed solving $D_Z W_{flux} = 0$ gives\footnote{See however \cite{CLZ} where  a non-negligible  correction to this result is computed.} $Z \sim \exp(-2\pi K/Mg_s)$.  However, as there will be further contributions to the superpotentional from the $\bDt$-brane, which we have not yet written down, we postpone the fixing of complex structure and dilaton, and keep the implicit dependence of $A_0$ on the moduli, defining this dependence as:
\be
e^{A_z(Z,\bar Z)} \equiv e^{4A_0 - 4u} \,. \label{E:Az}
\ee
At the end, it will be clear that the presence of the $\bD3$-brane in fact does not modify the stabilisation of $Z$ at exponentially small values, $Z \sim \exp(-2\pi K/Mg_s)$, and the hierarchy of scales $e^{A_{tip}} \sim Z^{1/3} \sim \exp(-2\pi K/3Mg_s)$. 
   
Dimensional reduction of the $\bDt$-brane action \eqref{E:SbD3quad} now gives the following result \cite{BDKVW} (compared to Eq. (3.18) and (A.14) in \cite{BDKVW} we change to the Einstein frame, include the volume fluctuations, use the unwarped metric and translate to two-component notation for the spinors\footnote{It will be clear from the context whether we are using 4-component Majorana spinors, 4-component complex Weyl spinors or 2-component spinors.}):
\bea
\calL^{\bDt}_{2-f} = \sqrt{-g} \, 2 T_3 e^{4A_0-12u} &&\left[e^{-A_0}\bar\lambda \bar\sigma^\mu \nabla_\mu \lambda +  e^{-A_0} \delta_{i\bar{i}}\bar{\psi}^{\bar{i}} \bar\sigma^\mu \nabla_\mu \psi^i \right. \nn \\
&& \left. + \frac12 e^{3A_0-6u}\bar{m}_{\bar{i}\bar{j}} \bar{\psi}^{\bar{i}} \bar{\psi}^{\bar{j}} + \frac12 e^{3A_0-6u} {m}_{ij} {\psi^i}{\psi^j} \right] \label{E:SbD3quad4D}
\eea
with
\be
m_{ij} = \epsilon_{j k l} \, e^{t}_{i} \, e^{\bar u}_{\bar k} \, e^{\bar v}_{\bar l} \delta^{k \bar k} \delta^{l \bar l} e^{\phi/2}\bar{G}^{ISD}_{t\bar{u}{\bar{v}}}\,. \label{E:mass}
\ee
  Provided that the background flux is of primitive (2,1) type, the triplet of worldvolume fermions, $\psi^i$, is massive, whereas the singlet, $\lambda$, remains massless.  The massless singlet, $\lambda$, is the goldstino of spontaneously broken 4D $\calN=1$ supersymmetry.  The symmetries in the setup imply that the completion of the action to higher order in the fermions would show the Volkov-Akulov nature of the theory, that is, non-linearly realised supersymmetry associated with spontaneous symmetry breaking.  Recall, moreover, that the bosonic part of an $\bDt$-brane action, in the GKP background, gives rise to an effective positive energy \cite{KKLMMT}:
\be
\V_{\bDt} = 2 T_3 e^{4 A_0-12u} \,. \label{E:bD3uplift}
\ee      
Note that, by itself, this potential energy would drive a runaway in the volume modulus, hence we will consider setups with additional non-perturbative effects, which ensure a stable compactification.     
Our task is now to work out the supersymmetric completion of the contributions \eqref{E:SbD3quad4D} and \eqref{E:bD3uplift} to the full 4D LEEFT, and write down the latter in the language of non-linear supergravity.

\section{Non-linear supergravity from $\bDt$-branes}

In this section we  work out the 4D LEEFT corresponding to a CY orientifold flux compactification including an $\bDt$-brane and non-perturbative effects, in terms of a supergravity theory with non-linearly realised supersymmetry.  We  first review the description of non-linear supersymmetry and the goldstino using constrained superfields, including their couplings to matter and supergravity.  Then we  work out the non-linear supergravity corresponding to the $\bDt$-brane in an ${\calN}=1$ supersymmetric GKP flux background -- including all bulk moduli fields.  As we have just seen, an $\bDt$-brane in a flux background gives rise to a runaway in the volume modulus\footnote{Without the $\bDt$-brane, a supersymmetric flux compactification leads to the well-known ''no-scale'' flat direction for the volume modulus.}, so we also include non-perturbative effects \`a la KKLT to stabilise this modulus in a metastable dS vacuum.  We  use three different formulations of the non-linear supergravity in terms of constrained superfields, where the uplift potential can be written either in terms of an F-term potential or a Fayet-Iliopoulos D-term, and show their equivalence.  

\subsection{Constrained superfields and their couplings to supergravity and matter}
So far, we have seen that the Volkov-Akulov theory provides a description for non-linear global supersymmetry.  The degrees of freedom correspond to the goldstini of spontaneously broken global supersymmetry, and the action is invariant under non-linear supersymmetry transformations acting on the goldstini.  The Volkov-Akulov theory can also be written with manifestly realised linear supersymmetry, using constrained superfields.  In fact, there are several different realisations with constrained superfields.

\paragraph{Constrained chiral superfield $X$}

The first such realisation was to place the goldstino within a constrained  chiral superfield\footnote{We write the chiral superfield as a function of its natural variables, $y^\mu = x^\mu + \i \theta \sigma^\mu \bar\theta$ and $\theta$.}, $X(y,\theta)=\phi^X(y) + \sqrt{2} \psi^X(y) \theta + F^X(y)\theta\theta$, obeying the following constraints\footnote{Recall that $-\frac14 \bar D^2 \bar\theta^2 = 1$.} \cite{Rocek}: 
\be
X^2 = 0 \quad \textrm{and} \quad -\frac14 X \bar{D}^2 \bar{X} = M^2 X \,. \label{E:X^2}
\ee
These constraints imply that the only independent degree of freedom in the chiral multiplet is the goldstino fermion, with the other components given by:
\be
\phi^X = \frac{\psi^X \psi^X}{F^X} \quad \textrm{and} \quad F^X = M^2 + \textrm{fermions} \,,
\ee
where the complete form of the second relation is given in \cite{KuzenkoTyler2} but is not important for our purposes.  Since $\langle F^X \rangle \neq 0$, supersymmetry is spontaneously broken.

  The action can be written either as a pure F-term or pure D-term (the second constraint implies that the F-term and D-term are equivalent): 
\be
\S = \int \d^4 x \int \d^2\theta \d^2\bar\theta X \bX \quad \textrm{or} \quad \S = M^2 \left(\int \d^4 x \int \d^2\theta X + h.c. \right) \,. \label{E:Xaction}
\ee
It was shown in \cite{Rocek} that this action is equivalent to the Volkov-Akulov one.

\paragraph{Constrained vector superfield $V$.}
Another possibility is to place the goldstino in a vector superfield:
\bea
V(x,\theta,\bar\theta) &=& C(x) + \i \theta \chi(x) - \i \bar\theta \bar\chi(x) + \frac{\i}{2}
\theta\theta \left[M(x) + \i N(x) \right] - \frac{\i}{2}\bar\theta \bar\theta\left[\bar M(x) - \i \bar N(x) \right] \nn \\
&&- \theta \sigma^\mu \bar\theta A_\mu(x) + \i \theta\theta\bar\theta\left[\bar\lambda(x) + \frac{\i}{2}\bar\sigma^\mu \partial_\mu \chi(x) \right] - \i \bar\theta\bar\theta\theta\left[\lambda(x) - \frac{\i}{2}\sigma^\mu \partial_\mu \bar\chi(x) \right] \nn\\
&&+ \frac12 \theta\theta\bar\theta\bar\theta\left[D(x)+ \frac12 \Box C(x) \right] \,, \label{E:Vcomps}
\eea 
 constrained as \cite{UlfRocek, SW}:
\bea
&&V^2 = 0,\, \nn \\
&&V D_A D_B V = 0\, \quad V D_A D_B D_C V = 0 \nn \\
&&\frac{1}{16} V D^\alpha \bar{D}^2 D_\alpha V = M^2 V \,\,\label{E:Vconstraints}
\eea
with $D_A = (\partial_a, D_\alpha, \bar D_{\dot{\alpha}})$.  These constraints fix all component fields -- including the gauge field -- in terms of the gaugino field $\lambda$, and remove the gauge symmetry.  

The Volkov-Akulov action, expressed in terms of a constrained vector superfield is \cite{UlfRocek, SW}:
\be
\S = -\frac12 M^2 \int \d^4x \int \d^2 \theta \d^2 \bar{\theta} V \quad \textrm{or}\quad \S =  \int \d^4x \int \d^2 \theta {\mathbb W}_\alpha {\mathbb W}^{\alpha} + h.c.\,,
\ee
with chiral field strength superfield ${\mathbb W}_\alpha = \frac14 \bar D^2 D_\alpha V$, and again, the derivative constraint ensures that the Fayet-Iliopoulos term and the kinetic term are equivalent

The constrained vector superfield and constrained chiral superfield are related by \cite{Bandos}:
\be
V = \frac{1}{M^2} \bX X \quad \textrm{and} \quad X = -\frac{M}{4} \bar{D}^2 V \,. \label{E:globVX}
\ee

\paragraph{Nilpotent chiral superfield, $S$.}  The constrained superfields $X$ and $V$ above have the goldstino as their only independent degree of freedom.  It is also possible to place the goldstino within a chiral superfield, $S$, that is constrained only by the nilpotency condition:
\be
S^2 = 0 \,. 
\ee
By dropping the derivative constraints, the complex auxiliary field, $F^S$ is restored to an independent (though non-dynamical) degree of freedom.  The Volkov-Akulov action in terms of the nilpotent chiral superfield, $S$, is given by:
\be
\S = \int \d^4 x \int \d^2\theta \d^2\bar\theta S \bS + M^2 \left(\int \d^4 x \int \d^2\theta S + h.c \right) \,.
\ee
 Note that a nilpotent vector superfield, constrained only by $V^2=0$, would carry -- in addition to the goldstino and real auxiliary field, $D$ -- the gauge field as an independent degree of freedom.  For a $\bDt$-brane on an O3$^-$-plane, the worldvolume gauge field is projected out, and therefore the degrees of freedom of $V^2=0$ do not correspond\footnote{Yet another possibility is to use the vector superfield recently introduced in \cite{Vsq, Vsq2}, constrained by $V^2=0$, $V D_A D_B V = 0$ and $V D_A D_B D_C V = 0$ (where $D_A = (\partial_a, D_\alpha, D^{\dot{\alpha}})$), which contains only the goldstino and auxiliary D-field as independent components and is thus the vector analogue of the nilpotent chiral superfield, $S$.} to those of the $\bDt$-brane on an O3$^-$-plane.  The superfield $S$ does carry the correct degrees of freedom, and is also convenient for coupling to matter and supergravity. 

\paragraph{Coupling to supergravity and matter.}
Having a superfield description of the goldstino makes its couplings to supergravity and matter straightforward.  In coupling to supergravity, the derivative constraints are modified by extending the flat superspace covariant derivatives to curved ones and the chiral projector to:
\be
\bar D^2 \rightarrow \bar{\mathcal D}^2 - 4R \,.
\ee
Derivative constraints are also modified by matter couplings.  These complications are avoided by taking the nilpotent field, $S$, for which the derivative constraint is discarded.  However, as already mentioned, in this case the auxiliary field, $F^S$ is then an independent degree of freedom, which has to be integrated out using its equation of motion.  This requires a complicated non-Gaussian integration  \cite{Kallosh:2016dcq, Bergshoeff:2016psz} after replacing $\phi^S = \psi^S \psi^S/F^S$.  In any case, the superspace action is described as usual by a real K\"ahler potential $K$, holomorphic superpotential $W$,  gauge kinetic functions $H_{AB}$ and Fayet-Iliopoulos D-terms:
\be
S = -\frac{3}{\kappa_4^2} \int \d^4x \d^2\theta \d^2\bar\theta E e^{-\frac13 \kappa_4^2 K(\Phi,\bar\Phi,V)}\, \Xi \bar \Xi - \frac{1}{\kappa_4^2} \int d^4x \d^2\theta {\mathcal E}\, \Xi^3 \left(W + H_{AB}(\Phi) \mathbb{W}^{A \alpha} \mathbb{W}^B_{\alpha}\right) + h.c.
\ee
where $\Xi$ is the so-called Weyl compensator field, a nowhere vanishing covariantly constant chiral superfield which renders the supergravity action invariant under scale and conformal transformations.  The bosonic part of the component action is just as in linear $\calN=1$ supergravity, and only the quartic and higher fermionic terms are different between the linear and non-linear supergravity \cite{PuredS, Kallosh:2015tea, Kallosh:2016dcq, Bergshoeff:2016psz}. 

\bigskip

Further possible constrained superfields which can carry the goldstino field are the spinor goldstino superfield \cite{SW} and a constrained complex linear superfield \cite{KuzenkoTyler}.  In addition to the goldstino superfields, there exist other constrained superfields containing matter fields with no superpartners.  We will encounter below a triplet of constrained chiral superfields, $Y^i$ ($i=1,2,3$), which are constrained by \cite{VW}:
\be
S Y^i = 0 \,.
\ee  
This constraint fixes the scalar component to be:
\be
\phi_i = \frac{\psi^S \psi^i}{F^S} - \frac{(\psi^S)^2}{2(F^S)^2} F^i \,.
\ee

\subsection{Non-linear supergravity for KKLT}

\label{S:NLsugraKKLT}

We are now ready to deduce the non-linear supergravity theory that describes, at low energies, a $\bDt$-brane in a supersymmetric GKP flux compactification with non-perturbative effects, {\it a.k.a.} a KKLT setup.  Let us review the fields present in the 4D LEEFT.  The flux background generically gives masses to the axio-dilaton, complex structure and worldvolume fermion triplet.  Usually, these fields are integrated out to leave only the K\"ahler moduli and goldstino (plus matter).  However, we  keep them, as their masses are suppressed with respect to the KK mass, and we want to know how the $\bDt$-brane couples to all the moduli and moreover consider constraints from invariance under modular transformations of the dilaton.  

With regards to the K\"ahler moduli, we have seen that at leading order, a $\bDt$-brane in a CY orientifold would give a runaway direction in the volume modulus.  We consider setups with non-perturbative racetrack effects arising from Euclidean D3-branes or gaugino condensation on wrapped D7-branes\footnote{Take care that gaugino condensation can provide a local source for supersymmetry breaking IASD flux $G_{(1,2)}$ \cite{Baumann:2010sx}, whereas we will assume  -- for simplicity -- supersymmetric background fluxes and  supersymmetric racetrack.  The affect of the $G_{(1,2)}$ fluxes in the 4D LEEFT can be parameterised by the usual superpotential of gaugino condensation \cite{Baumann:2010sx}.} which stabilise the volume modulus in a metastable dS vacuum, provided that the runaway contribution does not dominate the non-perturbative stabilisation \cite{racetrack}.  In that case, we can write down the 4D LEEFT\footnote{Note that the initial antibrane-flux setup -- before non-perturbative effects switch on -- has no static solution in 10D, as the positive energy density of the $\bDt$-brane would lead to decompactification.  Therefore, writing down a 4D LEEFT based around a static solution would be inconsistent.}.  Note that although we do not know the full 10D description of the non-perturbative effects, we do have a description of them in 4D.  Also, considering only classical and non-perturbative effects will be justified, as the superpotential is protected from perturbative corrections by non-renormalisation theorems, which we prove in Section \ref{S:NRtheorem}.

To summarise, we are assuming a mass hierarchy:
\be
 M^{w}_{s} \sim M_{\bcancel{susy}}  \gg  M^{w}_{kk} \sim  \Lambda \sim \Lambda_{np} \gg M_{\tau,Z}  \gg  M_{T}  \sim M_{3/2} \sim M_{goldstino} 
\ee
where $\Lambda$ is the cutoff for our 4D LEEFT and $\Lambda_{np}$ the energy scale at which non-perturbative effects kick in.  Note that there is no sgoldstino in our hierarchy.  In fact, the would-be superpartner of the goldstino in our setup is the worldvolume gauge boson, but the interplay between the $\bDt$-brane and the orientifold flux compactification leads to the non-linear realisation of supersymmetry already at the warped string scale,  $M^{w}_{s}$, and the goldstino and gauge boson should lie in separate constrained supermultiplets as in \cite{VW, Kallosh:2016aep}.  As we have seen, the decompactification limit does not restore supersymmetry -- the $\bDt$-brane in the flat orientifold still has only non-linear supersymmetry.  Supersymmetry is only restored after the process of antibrane-flux annihilation, where the supersymmetry breaking sectors of the $\bDt$-brane are replaced by supersymmetry preserving degrees of freedom on a stack of D3-branes.  The goldstino degree of freedom on the antibrane can be written as a constrained $\calN=1$ superfield, and we now want to identify the goldstino superfield and its couplings to the bulk moduli.

The leading order action governing the 4D LEEFT can be deduced using a combination of dimensional reduction, non-linear supersymmetry and modular invariance. 

\subsubsection{Modular Invariance}
\label{S:modinv}
A fundamental feature of string theory is modular invariance or S-duality.  In type IIB string theory, this can be seen as descending from the modular invariance of the torus in the elliptic fibration that relates type IIB to F-theory.  Indeed, the 10D axio-dilaton\footnote{For convenience of notation, we now rotate the axio-dilaton given in Section \ref{S:setup} by $-\i$.},  $\tau = e^{-\phi}- \i \, C_0$, parameterises an $SL(2,\R)/U(1)$ coset space.  The type IIB supergravity action in the Einstein frame \eqref{E:IIBaction} is manifestly invariant under the $SL(2,\R)$ transformations, where it will be useful to see how the following combinations of fields transform:
\bea
&&\tau\rightarrow \frac{a\tau - \i b}{\i c\tau+d}, \quad G_{(3)} \rightarrow \frac{G_{(3)}}{\i c\tau + d}, \quad e^{\phi/2} G_{(3)} \rightarrow e^{-2\i \delta} e^{\phi/2} G_{(3)}, \quad F_{(5)} \rightarrow F_{(5)} \nn \label{E:SL2Zxfmn} \\ \nonumber \\
&&
\Psi_M \rightarrow e^{-\i \delta}\Psi_M, \quad \lambda \rightarrow e^{-\i 3 \delta} \lambda\,,
\eea
with $a,b,c,d \in \R$ and $ad-bc=1$ and the phase 
\be
e^{\i\delta } = \frac{(\i c \tau + d)^\frac12}{|\i c \tau +d|^\frac12}\,.
\ee
    This modular invariance of the type IIB classical action is broken at the perturbative level and restored to $SL(2,\Z)$ by non-perturbative effects.  
		
We  begin by deriving the classical contributions to the 4D LEEFT action, which inherit the perturbative $SL(2,\R)$ invariance.  Describing the 10D gravitino that survives the orientifolding as a 10D MW spinor, its dimensional reduction gives\footnote{Here, $c.c.$ stands for charge conjugation, $\Psi^C \equiv C_{10} \bar \Psi^T$.  Writing the decomposition of the 10D charge conjugation matrix is $C_{10} = C_4 \otimes C_6$, we have $c.c. = -C_{4} \gamma^0 (\psi_\mu^+)^* \otimes C_6 (\zeta_-)^*$,  so $\psi_\mu^+ \otimes \zeta_- \rightarrow e^{-\i \gamma_5 \delta} \psi_\mu^+ \otimes \zeta_-$ and $c.c. \rightarrow e^{-\i \gamma_5 \delta} c.c.$ under modular transformation.}:
\be
\Psi_\mu = \psi_\mu^+ \otimes \zeta_- + c.c.\,,
\ee
with $\zeta_-$ the 6D nowhere vanishing covariantly constant (w.r.t. the unwarped metric) spinor with negative chirality and $\psi_\mu^+$ the 4D massless gravitino in complex Weyl notation with positive chirality. Converting to the usual 4D Majorana gravitino, the latter transforms as:
\be
\psi_\mu \rightarrow e^{-\i \delta \gamma_5 }\psi_\mu
\ee
from which we can deduce the transformation of the supersymmetry Killing spinor and the fermionic superspace coordinate:
	\be
	\epsilon \rightarrow  e^{-\i \gamma_5 \delta}\epsilon \quad \textrm{and} \quad \theta \rightarrow  e^{-\i \gamma_5 \delta} \theta \,.
	\ee
	
In the absence of the antibrane, $\calN=1$ supersymmetry is preserved.  The 4D axio-dilaton and volume modulus fall into ${\mathcal N}=1$ chiral multiplets, whose complex scalar components are given by:
\be
\tau = e^{-\phi} - \i \, C_0 \quad \textrm{and} \quad T = e^{4u} - \i \, b
\ee
where, recall that $ds^2 = e^{2A-6u(x)} g_{\mu\nu} dx^\mu dx^\nu + e^{-2A+2u(x)} g_{mn} dy^m dy^n$ and $b$ is the universal axion descending from the 10D self-dual 4-form $C_{(4)}$, $C_{(4)} = a_{(2)} \wedge \tilde{J}$ and $\d a_{(2)} = e^{-8u} \tilde{\star} \d b$.  The low energy effective field theory is described by an ${\cal N}=1$ 4D supergravity action, with K\"ahler potential and superpotential taking the form (as usual, chiral superfields are labelled by their lowest components):
\be
\kappa_4^{2} \,K = - \ln(\tau + \bar \tau) - 3 \ln(T+\bar T) -  \ln\left(-\i\int_M \Omega \wedge \bar\Omega\right)
\ee
and 
\be
 W =  \int G_3 \wedge \Omega \,,
\ee
where $\kappa_4^2 = \kappa_{10}^2 /{\langle{\cal V}^w\rangle}$, with ${\cal V}^w$ is the warped 6D volume.  Moreover, $\Omega$ is the holomorphic (3,0)-form of the CY, so that  the last term in the K\"ahler potential gives the dependence on the complex structure moduli.

The K\"ahler and superpotential are invariant under the 4D modular transformation:
\be
\tau \rightarrow \frac{a\tau - \i b}{\i c\tau+d}\,, \label{E:SL2Zxfmn4D}
\ee
up to a K\"ahler transformation\footnote{Recall that the transformation \eqref{E:SL2Zxfmn4D} implies that the dilaton chiral supermultiplet $\tau(y,\theta)$ transforms as:
\be
\tau(y,\theta) \rightarrow \frac{a \tau(y,\theta') - \i b}{\i c\tau(y,\theta') + d} \,.
\ee
This induces a K\"ahler transformation in the K\"ahler potential, $K = -\ln(\tau + \btau)$, from which we can write down how the susy parameter, $\theta_\alpha$, transforms:
\be
K \rightarrow K + F + \bar{F}, \quad \textrm{with} \quad F = \ln(\i c\tau + d), \quad \Rightarrow \quad \theta_\alpha \rightarrow \theta_\alpha'= \frac{|\i c\tau + d|^\frac12}{(\i c\tau + d)^\frac12} \theta_\alpha\,. \label{E:modKxfmn}
\ee
 The K\"ahler transformation is also accompanied by Weyl rotations of the spinors as follows  \cite{WessBagger}.  The gravitino transforms as $\psi_\mu \rightarrow \exp(-\frac{\i}{2}(\Im F))\psi_\mu$, as does the spinor in a vector superfield, $\lambda \rightarrow \exp(-\frac{\i}{2}(\Im F))\lambda$.  Spinors of modular weight 0 transform as $\psi \rightarrow \exp(\frac{\i}{2}(\Im F))\psi$.}
\be 
K \rightarrow K + \ln|\i c \tau + d|^2 \quad \text{and} \quad W \rightarrow W/(\i c \tau + d)
\ee
so that the combination $G = K + \ln|W|^2$ remains invariant.

The modular symmetry persists in the presence of probe D3/$\bDt$-branes, which are each self-dual under the $SL(2,\R)$ transformation \cite{D3dual1,D3dual2,D3dual3}.  Requiring that the coupling  \eqref{E:SbD3quad} between the 10D MW spinor, $\Theta$, and the field strength, $G_{(3)}$, in the Einstein frame, is modular invariant allows us to deduce the modular transformation property of the worldvolume fermion \cite{Grana}.  For the $\bDt$-brane (see appendix):  
\be
\Theta \rightarrow e^{\i (\gamma_5 \otimes \Id) \delta}  \Theta \,.
\ee
Dimensional reduction of the worldvolume fermion on the $\bDt$-brane gives:
\be
\Theta = \lambda_-\otimes \zeta_- + \psi_-^i \otimes \zeta^i + c.c.
\ee
where $\lambda_-$ and $\psi_-^i$ are 4D complex Weyl spinors with negative chirality, $\zeta_i = \frac{1}{6||\Omega||}\Omega_{ijk} \gamma^{jk} \zeta_-$ and $||\Omega||^2=\frac{1}{3!}\Omega_{ijk}\bar\Omega^{ijk}$.  The 4D singlet and triplet transform under the modular symmetry as (converting to 4D Majorana spinors): 
\be
 \lambda \rightarrow  e^{-\i\gamma_5\delta} \lambda
\ee
and similarly for the triplet of worldvolume fermions:
\be
\psi^i \rightarrow  e^{-\i\gamma_5\delta} \psi^i \quad (i=1,2,3).
\ee
These imply, as they must, that the 4D action descending from the $\bDt$-brane is modular invariant. 

  Notice that the classical action also enjoys an accidental $SL(2,\R)$ symmetry acting on the K\"ahler modulus, whose Peccei-Quinn subgroup descends from the 10D RR-gauge invariance.

\subsubsection{Nilpotent chiral superfield and modular invariance}
We now write down the non-linear supergravity with nilpotent chiral superfield, corresponding to a KKLT model.
We have seen that the $\bDt$-brane introduces four new fermionic degrees of freedom corresponding to a goldstino and fermion triplet.  Refs. \cite{KW, BDKVW} propose that the goldstino lies in the nilpotential chiral superfield, satisfying $S^2 = 0$, and Ref. \cite{VW} argues that the fermion triplet lies in three constrained chiral supermultiplets, satisfying $S Y^i = 0$.  We now consider the same constrained chiral superfields, together with all the light bulk moduli.

\paragraph{GKP flux compactification with an $\bDt$-brane.}
Let us start with the contributions to the 4D LEEFT action that descend from the CY orientifold compactification in the presence of background fluxes and a probe $\bDt$-brane.   These contributions stabilise the complex structure and dilaton with a positive vacuum energy, but -- if they were the only contribution -- would leave a runaway direction in the volume modulus.  A candidate K\"ahler potential and superpotential is\footnote{Our conventions for the mass dimensions are $[\theta]=-\frac12$ (so $[\d\theta] = \frac12$ and $[\mathcal D_\alpha = \frac12]$), $[S] = 1$ and $[V] = 0$.}:
\bea
\kappa_4^2 K &=& - \ln(\tau + \bar \tau) - 3 \ln(T+\bar T) - \ln\left(-\i\int_M \Omega \wedge \bar\Omega\right) \nn \\
&&+ K_{S\bar S}(\tau, \bar \tau, T, \bar T, Z, \bar Z){S\bar S}  +  K_{Y\bar Y}(\tau, \bar \tau, T, \bar T, Z, \bar Z){\delta_{i\bar{i}} Y^i \bar Y^{\bar i}} 
\eea
and
\be
W = \int G_3 \wedge \Omega + M(\tau,T,Z)^2 S+ h_{ij}(\tau,T,Z) Y^i Y^j 
\ee
where $K_{S\bar S}(\tau, \bar \tau, T, \bar T, Z, \bar Z),  K_{Y \bar Y}(\tau, \bar \tau, T, \bar T, Z, \bar Z), M(\tau,T,Z)$ and $h_{ij}(\tau,T,Z)$ are functions of the moduli -- possibly constant -- to be identified.  We have neglected possible contributions $K_{YY}Y^i Y^j +  h.c.$ to the K\"ahler potential, as such terms would contribute to the component action only at higher order in the fermions, and thus cannot be fixed to the quadratic order we are working in, see \cite{Louis} for analogous terms for a D3-brane in a flux background. Also, the most general action would include a linear term in the superpotential, $g_i(\tau,T,Z) Y^i$, but this term is absent for our choice of background fluxes.

Let us first assume an unfluxed, unwarped CY orientifold with no complex structure.  We first write down a candidate real K\"ahler potential and holomorphic superpotential that would satisfy modular invariance and match the uplift term in the scalar potential and fermion mass terms from the $\bDt$-brane.  Then we match the fermion kinetic terms, and  -- as a consistency check -- ensure that the superfields derived via this matching have the assumed modular transformation properties.   The result is the following:
\bea
\kappa_4^2 K &=& - \ln(\tau + \bar \tau) - 3 \ln\left(T+\bar T - \kappa_4^2 \frac{(T+\bar T)}{3(\tau+ \bar \tau)} S \bar S  - \kappa_4^2 \frac13 (T+\bar T) \delta_{i\bar{i}} Y^i \bar Y^{\bar i}\right) \,\,,
\eea
and 
\be
W = M^2 S \quad \textrm{where} \quad M^2 = \sqrt{2 T_3}   \,, 
\ee 
with
\be
\psi^{S} = e^{-\phi/2} e^{-6u} M^2 \lambda + \dots\,\, \text{ and } \,\, \,\,\psi_{Y}^i = M^2 e^{-6u} \gamma^0 \psi^{i} + \dots\,,
\ee
where here and below we drop higher order fermion terms in the fermion field redefinitions.  Indeed, from the $SL(2,\R)_\tau$ modular transformation properties of $\lambda$, $\psi^i$ and $\theta$, we can deduce that the superfields $S$ and $Y^i$ transform as\footnote{Before fluxes are included, $\psi_{Y}^i = M^2 e^{-6u} \psi^{i} + \dots$ and $Y^i \rightarrow e^{-2\i \delta}Y^i$ would also seem a consistent choice, however, writing the flux-dependent mass term as a holomorphic superpotential term fixes $Y^i$ as in the main text.}:
\be
 S\rightarrow \frac{S}{\i c\tau+d} \quad \text{and} \quad Y^i \rightarrow Y^i \,.
\ee
Note also that $S$ and $Y^i$ then have ``modular weight'' -1 or 0 under the $SL(2,\R)_T$ symmetry, depending on whether the antibrane is in the highly warped or unwarped region of the CY.

  The corresponding kinetic term for the goldstino and fermion triplet is:
\be
\calL^{\bDt}_{2-f} = \sqrt{-g} \, \left[\frac{1}{(\tau+ \bar \tau)}  \bpsi^S \bar\sigma^\mu \nabla_\mu \psi^S +  \, \delta_{ij} \bpsi_Y^i \bar\sigma^\mu \nabla_\mu \psi_Y^j\right]
\ee
and the contribution to the scalar potential is
\be
\V_{\bDt} = \frac{M^4}{(T+\bar T)^{3}} \,.
\ee
The $K$ and $W$ also lead to a number of other couplings which are higher order than the approximations we used in the dimensional reduction, but whose presence are guaranteed by supersymmetry.  As well as higher than quadratic order in the worldvolume fermions, we also thus derive couplings between bulk and brane fields (e.g.~via the supercovariant derivatives of the fermions), which we neglected when evaluating the probe brane action in a background supergravity configuration.

Note that, although the $\bDt$-brane breaks supersymmetry, the Lagrangian gravitino mass parameter $m_{3/2} = \kappa_4^2 \vline e^{\kappa_4^2 K/2}W \vline$  vanishes since $W=0$ in the background.  We will return to the gravitino mass once we have incorporated non-perturbative contributions and stabilised the volume modulus in a genuine dS vacuum, but see also \cite{KalloshVP, FerraraVP, superHiggs} for a discussion of the gravitino mass in dS space.

Now let us assume a warped, fluxed CY orientifold with a single complex structure modulus.  Before including the $\bDt$-brane, the K\"ahler potential is:
\be
\kappa_4^{2} K = - \ln(\tau+ \bar \tau)- 3 \ln(T+\bar T) - 3 \ln(Z+\bar Z) 
\ee
After adding the $\bDt$-brane, we need that the uplift potential be independent of the complex structure.  Since the superpotential is holomorphic, the only possibility is:
\bea
\kappa_4^{2}\,K &=& -  \ln(\tau + \bar \tau) - 3 \ln(Z+\bar Z) \nn \\
&&- 3  \ln\left(T+\bar T - \kappa_4^2 \frac{e^{-4A_0}(T+\bar T)}{3(\tau+ \bar \tau)(Z + \bar Z)^3}  S \bar S - \kappa_4^2 \frac{e^{-4A_0} (T+\bar T)}{3(Z+ \bar Z)^3} \delta_{i\bar{i}} Y^i \bar Y^{\bar i}\right)  
\eea
and
\be
W = \int G_{(3)} \wedge \Omega + M^2 S + h_{ij}(\tau) Y^i Y^j \quad \textrm{with} \quad M^2 = \sqrt{2 T_3}  \,,
\ee 
where
\be
\psi^{S} = e^{-\phi/2} e^{-6u + 7A_0/2} (Z+\bar Z)^{3/2} M^2  \lambda + \dots \text{ and }\,\, \psi_{Y}^i = M^2 e^{-6u + 7A_0/2} (Z+\bar Z)^{3/2}\gamma^0 \psi^{i} + \dots 
\ee
and $h_{ij}(\tau)$ is a function of the dilaton determined from \eqref{E:mass} to be:
\be
h_{ij}(\tau) = (Z+\bar Z)^{-\frac32} \epsilon_{\bar j \bar k \bar l} \, e^{\bar t}_{\bar i} \, e^{u}_{k} \, e^{v}_{l} \delta^{k \bar k} \delta^{l \bar l} \delta_i^{\bar i}\delta_j^{\bar j} {G}^{ISD}_{\bar t {u}{v}} \,,\label{E:hij}
\ee
recalling that all quantities are with respect to the unwarped metric, $g_{mn}$ in \eqref{E:metricwv}.  
Notice that as $h_{ij}(\tau)$ is linear in the flux $G_{(3)}$, it has modular weight -1, as required by modular invariance. Moreover, similarly to the case of a D3-brane in a flux background \cite{Louis}, all the non-holomorphic dependence on $Z$ in the right-hand side of \eqref{E:hij} must cancel, yielding a holomorphic superpotential independent of $Z$.  We have checked this for the simple toroidal orientifold outlined in Appendix B of \cite{BDKVW}.
  
 The classical plus antibrane contributions to the scalar potential are then:
\be
\V_{cl} = \V_{flux} + \frac{M^4 e^{4A_0}}{(T+\bar T)^{3}} =  \V_{flux} + \frac{M^4 e^{4A_z(Z,\bar Z)}}{(T+\bar T)^{2}}\,.
\ee
where the second equality is for the antibrane placed in the highly warped region at the tip of the throat (where $e^{4A_0} = e^{4u + A_z(Z, \bar Z)}$).  Note that we have chosen a background with $\langle W \rangle = 0$, $\langle D_\tau W \rangle = 0$, $\langle D_Z W \rangle = 0$ and $\langle D_T W \rangle = 0$, and supersymmetry is broken only by $\langle D_S W \rangle \neq  0$.  Therefore, the potential so far is semi-positive definite, stabilises the complex structure and dilaton, but as discussed, leaves the volume modulus as a runaway to be stabilised by non-perturbative effects.

For a general CY, we have:
\bea
\kappa_4^{2}\, K &=& -  \ln(\tau + \bar \tau)-  \ln f(Z, \bar Z) \nn \\
&& -3  \ln\left(T+\bar T - \kappa_4^2 \frac{e^{-4A_0}(T+\bar T) }{3(\tau+ \bar \tau)f(Z,\bar Z)} S \bS  -  \kappa_4^2 \frac{e^{-4A_0}(T+\bar T)}{3f(Z,\bar Z)} \delta_{i\bar{i}} Y^i \bar Y^{\bar i}\right)  \label{E:Kfinal}
\eea
and
\be W = \int G_3 \wedge \Omega + M^2 S + h_{ij}(\tau) Y^i Y^j\,, \label{E:KWfinal}
\ee
with
\be
h_{ij}(\tau) = f(Z,\bar Z)^{-\frac12} \epsilon_{\bar j \bar k \bar l} \, e^{\bar t}_{\bar i} \, e^{u}_{k} \, e^{v}_{l} \delta^{k \bar k} \delta^{l \bar l} \delta_i^{\bar i}\delta_j^{\bar j} {G}^{ISD}_{\bar t {u}{v}}
\ee
and
\be
\psi^{S} = e^{-\phi/2} e^{-6u + 7A_0/2} f(Z,\bar Z)^{1/2} M^2  \lambda + \dots \text{ , }\,\,\,\, \psi_{Y}^i = M^2 e^{-6u + 7A_0/2} f(Z,\bar Z)^{1/2} \gamma^0 \psi^{i} + \dots \,.
\ee
Note that the above relations -- together with the 4D supergravity supersymmetry transformations $\delta \psi^S = - \sqrt{2} F^S \epsilon + \dots$ -- imply that the normalised Volkov-Akulov brane fields $\lambda_{VA} = e^{-A_0/2}\lambda$, $\psi_{VA}^i = e^{-A_0/2}\psi^i$ (see \eqref{E:SbD3quad4D}) have the expected non-linear supersymmetry transformation (working to lowest order in the fermions):
\bea
&&\delta \lambda_{VA}  = \epsilon + \mathcal{O}(fermion^2)\\
&&\delta \psi^i_{VA} = \mathcal{O}(fermion^2) \,.
\eea
We also recover, as is necessary, the expected linearly supersymmetric 4D $\calN=1$ theory when the $\bDt$-brane tension is taken to zero.

\paragraph{Non-perturbative effects, the dS vacuum and the super-Higgs mechanism.}  The classical plus antibrane contributions to the scalar potential discussed so far, although positive semi-definite, give a runaway in the volume modulus, $T$.  However, non-perturbative contributions will also come into play, and may stabilise the volume modulus leading to a metastable dS vacuum, if they are not too small.  For example, one can consider a racetrack combination of gaugino condensates from wrapped D7-branes and/or Euclidean D-branes.  The final 4D non-linear $\calN=1$ LEEFT is then given by:
\bea
\kappa_4^{-2} \,K &=& -  \ln(\tau + \bar \tau)-   \ln f(Z, \bar Z) \nn \\
&& -3  \ln\left(T+\bar T - \kappa_4^2 \frac{e^{-4A_0}(T+\bar T) }{3(\tau+ \bar \tau)f(Z,\bar Z)} S \bS  -  \kappa_4^2 \frac{e^{-4A_0}(T+\bar T)}{3 f(Z,\bar Z)} \delta_{i\bar{i}} Y^i \bar Y^{\bar i}\right)
\eea
and
\bea
&& W = \int G_3 \wedge \Omega + M^2 S + h_{ij}(\tau) Y^i Y^j + A e^{-aT} + B e^{-bT}\label{E:KWfinalNP}
\eea
Note that the non-perturbative $SL(2,\Z)$ modular invariance suggests that the racetrack contributions to $W$ carry some flux or matter dependence, which imparts a modular weight -1, through the coefficients $A$, $B$.  The gravitino mass is now $m_{3/2} = \kappa_4^2 |e^{\kappa_4^2 K/2} W| = \kappa_4^2 |e^{\kappa_4^2 K/2} W_{racetrack}|$.

\subsubsection{Modular invariance with constrained superfields $X$ or $V$}
The non-linear supergravity describing the $\bDt$-brane in a GKP flux background at low energies can also be written using a constrained chiral superfield, $X$, satisfying\footnote{The Weyl weights are as follows: $[\Xi]=-1$, $[\bar{\cal D}_{\dot{\alpha}}] = -\frac12$, $[{\cal D}_\alpha]=-\frac12$, $[X]=0$ and $[V]=0$.} \cite{Bandos}:
\be
X^2 = 0 \quad \textrm{and} \quad M^2 X \, \Xi^3 = -\frac14 {\mathcal G}X\left(\bar{{\cal D}}^2 - 4R \right) {\mathcal F} \bar X  \Xi\, \bar \Xi\,, \label{E:DXconstraint}
\ee
where  ${\mathcal F}$, ${\mathcal G}$ are, respectively, composite real and covariantly chiral superfields, functions of the moduli fields to be determined in order to set the supersymmetry breaking auxiliary field $F^X$ as required. 

We fix the conformal gauge to the Einstein frame (see e.g. \cite{Butter:2009cp}):
\be
\Xi = e^{\kappa_4^2 K_{moduli}/6} = (\tau+\bar \tau)^{-\frac16} \, f(Z,\bar Z)^{-\frac16} \, (T+\bar T)^{-\frac12} \,.
\ee

The scalar potential is given in terms of $F^X$ as\footnote{See e.g. equation (6.26) of \cite{Binetruy} for the scalar potential in terms of the supergravity auxiliary fields.  We have integrated out the auxiliary field $M$ of the supergravity multiplet.}:
\be
V_F = -F^X K_{X\bar X} F^{\bar X} -  e^{\kappa_4^2 K/2}\left(F^X D_X W + F^{\bar X}D_{\bar X} \bar W \right) - 3\kappa_4^2 e^{\kappa_4^2 K} |W|^2\,.
\ee
with $F^X$ determined by the derivative constraint in \eqref{E:DXconstraint}.
We can then write the action, for example, as:
\bea
\kappa_4^{2}\, K &=& -  \ln(\tau + \bar \tau)-   \ln f(Z, \bar Z) \nn \\
&& -3  \ln\left(T+\bar T - \kappa_4^2 \frac{e^{-4A_0}(T+\bar T) }{3(\tau+ \bar \tau)f(Z,\bar Z)} X \bX  -  \kappa_4^2 \frac{e^{-4A_0}(T+\bar T)}{3f(Z, \bar Z)} \delta_{i\bar{i}} Y^i \bar Y^{\bar i}\right) \label{E:KDX}
\eea
and
\be
W = \int G_3 \wedge \Omega + M^2 X+ h_{ij}(\tau) Y^i Y^i \label{E:WDX}
\ee
where 
\be
\psi^{X} = e^{-\phi/2} e^{-6u+7A_0/2} f(Z,\bar Z)^{1/2} M^2  \lambda + \dots, \quad \psi_{Y}^i = M^2 e^{-6u+7A_0/2} f(Z,\bar Z)^{1/2} \gamma^0 \psi^{i} + \dots
\ee
and the derivative constraint is chosen in order to  match the uplift potential with an F-term potential:
\be
V_F = \frac{M^4 e^{4A_0}}{(T+\bar T)^3} \quad \Rightarrow \quad F^X = -\frac{(\tau + \bar\tau)^\frac12f(Z,\bar Z)^{\frac12}}{(T + \bar T)^\frac32} e^{4A_0} M^2 + \textrm{fermions} \label{E:FX}
\ee
so:
\be
{\mathcal G} = 1 \quad \textrm{and} \quad {\mathcal F} = -(\tau + \bar\tau)^{-2/3}f(Z,\bar Z)^{-2/3} (T+\bar T)e^{-4A_0}  + \textrm{fermions} \,.\label{E:mathcalF}
\ee
Note that the derivative constraint in \eqref{E:DXconstraint} also implies that the superspace action can moreover be equivalently written either in terms of a K\"ahler contribution or a superpotential contribution from $X$.

Alternatively, a constrained vector superfield, $V$, may be used, satisfying \cite{Bandos}:
\be
V^2 = 0 \quad \textrm{and} \quad \frac{1}{16} {\mathcal H} V {\cal D}^\alpha \left(\bar{{\cal D}}^2 - 4R \right) {\cal D}_\alpha {\mathcal I} V = M^2 V \Xi \bar \Xi \,, \label{E:Vconstraintsugra}
\ee
where recall that the derivative constraint removes the gauge field and gauge symmetry, and fixes the auxiliary D-term. Again, ${\mathcal H}$, ${\mathcal I}$ are functions of the moduli fields to be determined in order to set the supersymmetry breaking auxiliary field $D$ as required. The scalar potential is given by the following D-term contributions:
\be
V_D  = -\frac{\Re(H)}{2} D^2 - D^a \left(K_V|_{V=0}\right) \,,
\ee
where $H$ is the gauge kinetic function.  One way to write the action is then:
\bea
  \kappa_4^{2}\,K &=& -\ln(\tau + \bar \tau)- \ln f(Z, \bar Z)
 -3  \ln\left(T+\bar T -  \kappa_4^2 \frac{e^{-4A_0}(T+\bar T)}{3 f(Z,\bar Z)} \delta_{i\bar{i}} Y^i \bar Y^{\bar i} - \xi V\right)\,,   \nn \\ \\
&&H=-1 \quad \text{and} \quad W = \int G_3 \wedge \Omega  + h_{ij}(\tau) Y^i Y^i \label{E:Vaction}
\eea
where: 
\be
\lambda_{V} = e^{-6u+3A_0/2} M^2 \gamma^0 \lambda + \dots, \quad \psi_{Y}^i = M^2 e^{-6u +7A_0/2}  f(Z,\bar Z)^{1/2} \gamma^0 \psi^{i} + \dots \,.
\ee
The derivative constraint is chosen in order to  match the uplift potential with a D-term potential:
\be
D = -\frac{2 M^2 e^{2A_0}}{(T + \bar T)^\frac32} + \textrm{fermions}  \quad \textrm{and} \quad \xi = -\frac{\kappa_4^2 M^2}{6}  \frac{e^{2A_0}}{(T+\bar T)^\frac12}\label{E:Dxi}
\ee
so:
\be
{\mathcal H} = 1 \quad \textrm{and} \quad {\mathcal I} = -\frac12 e^{-2A_0}(\tau + \bar \tau)^{-\frac13} f(Z, \bar Z)^{-\frac13}(T+\bar T)^{\frac12} + \textrm{fermions} 
\ee
 Now the derivative constraint is such that the superspace action could also be written equivalently as either a kinetic term for the vector supermultiplet or a Fayet-Iliopoulos term.  Note that although $V$ enters as a field dependent FI-term, it is not associated with a gauged $U(1)$ shift symmetry or St\"uckelberg coupling, as there is no dynamical gauge field.

Finally, note that the above action is modular invariant, with $V$ modular invariant, as it must be given that the hermitian conjugate components in the $\theta$ expansion of the superfield \eqref{E:Vcomps} would otherwise transform with opposite weights.

\paragraph{Pure D-term uplifting and the super-Higgs mechanism.}
The appearance of a D-term uplift simultaneously with a trivial F-term deserves some further comment.  Indeed, it is usually said that D-terms are proportional to F-terms, and can only be non-vanishing if the latter are non-vanishing (see e.g. \cite{VZ}).  The auxiliary field $D$ is called the Killing potential, and is the real solution to the complex Killing equation:
\be
(t^a \phi)^i = -\i G^{i\bar{k}} \frac{\pd D^a}{\pd\bar\phi^{\bar{k}}} \quad \Rightarrow \quad D^a = \i G_i (t^a \phi)^i = \i K_i (t^a \phi)^i + \i \frac{W_i}{W} (t^a \phi)^i \quad \Rightarrow \quad e^{G/2} D^a = F_i X^i \label{E:killingeqn}
\ee
with $t^a$ the generator of the would-be gauged isometry on the K\"ahler manifold spanned by the $\phi^i$, and $F_i = e^{G/2} G_i$.  However, for the constrained vector field, there is no gauged isometry and \eqref{E:killingeqn} simply does not arise.  Similar to the F-term uplifting associated with the nilpotent chiral superfield, the gravitino mass is given by $m_{3/2} = \kappa_4^2 |e^{\kappa_4^2 K/2} W|=\kappa_4^2 |e^{\kappa_4^2 K/2} W_{racetrack}|$.  Note that the non-perturbative effects are not subject to a gauge invariance associated with the D-term, as there is no associated gauged symmetry, but they must be modular invariant.

\subsubsection{Equivalence between the F-term and FI D-term uplift}

The 4D LEEFT in terms of the constrained chiral supermultiplet, $X$,  and the constrained vector supermultiplet, $V$, are equivalent.  In the absence of matter, the relations between the two fields are given by \eqref{E:globVX}.  In the presence of matter, these relations are modified.  Starting from the action in terms of $X$, \eqref{E:KDX} and \eqref{E:WDX}, we can obtain the action in terms of $V$,  \eqref{E:Vaction}, using the following relations.  Firstly\footnote{Note that \eqref{E:XtoVsugra} can be used to relate the auxiliary field $F^X$ and the auxiliary field $D$, and that this relation agrees with that obtained directly from  \eqref{E:FX} and \eqref{E:Dxi}.}:
\bea
&&X \bar X  = M^2 \mathcal{J} V \quad \textrm{with} \quad \mathcal{J} = -\frac{e^{6A_0}(\tau+\bar\tau)f(Z, \bar Z)}{2(T+\bar T)^\frac32} + \textrm{fermions}  \,. \label{E:XtoVsugra}
\eea
Secondly, the inverse relation between $X$ and $V$, which can be derived using \eqref{E:XtoVsugra} and \eqref{E:DXconstraint}:
\bea
M^2 X\, \Xi^3 = -\frac14 (\bar{\cal D}^2 - 4R) \mathcal{F} \mathcal{J} M^2 V \Xi \,\bar \Xi\,.  \label{E:VtoXsugra}
\eea
The superpotential term $M^2X$ in \eqref{E:WDX}, becomes a gauge kinetic term in \eqref{E:Vaction} with the help of \eqref{E:VtoXsugra} and \eqref{E:Vconstraintsugra} (and recall that the kinetic term for the vector supermultiplet is moreover equivalent to an FI-term).

\section{Non-Renormalisation Theorem}
\label{S:NRtheorem}

In the previous section, we have presented the non-linear supergravity that constitutes the 4D non-linear $\calN=1$ LEEFT describing a flux orientifold compactification of type IIB string theory in the presence of a probe $\bDt$-brane (placed on an O3$^-$-plane) and non-perturbative effects.  A generic non-linear supergravity action would be subject to quantum corrections, and in fact, be highly non-renormalizable.  However, as EFTs, the non-linear supergravity theories in which we are interested are valid only up to a cutoff given by the KK scale, with non-renormalisable interactions suppressed by inverse powers of the cutoff scale.  Moreover, as we  now show, non-linear supergravity theories arising from string theory have an enhanced protection from quantum corrections. This is due to a straightforward extension of the well-known non-renormalisation theorems that apply to the 4D EFTs describing linearly realised $\calN=1$ string compactifications.

The classic string theory non-renormalisation theorems \cite{Witten:1985bz,Dine:1986vd} are based on the holomorphicity of the superpotential $W$ and the gauge kinetic function $f$ and the invariance under certain global symmetries -- the perturbative Peccei-Quinn (PQ) symmetries $\tau \rightarrow \tau + \i \, const$ and $T \rightarrow T + \i \, const$.  These symmetries preclude the superpotential, $W$, and gauge kinetic function, $f$, from depending perturbatively on the axionic parts of the dilaton and volume chiral multiplets.  Holomorphicity of the $W$ and $f$ then also precludes them from depending on the string coupling or volume modulus, and therefore $W$ receives no corrections to all finite orders in the string-loop and $\alpha'$ expansions, and $f$ receives corrections at most at one-loop.  For flux compactifications, the discussion is a bit more involved, as the tree-level superpotential does depend explicitly on the dilaton via the GVW flux contribution.  Ref. \cite{BEQ} extended the textbook arguments to this case, and we  now do the same for flux compactifications in the presence of an $\bDt$-brane with non-linearly realised supersymmetry, following \cite{BEQ}.  As well as the PQ-symmetries used in string non-renormalisation theorems, R-symmetry as used to prove the non-renormalisation of global supersymmetry theories \cite{Seiberg:1993vc}  also play an important role.

Recall that the 4D non-linearly  realised ${\mathcal N}=1$ supergravity theory is parameterised - similarly to the case of linearly realised ${\mathcal N}=1$ supersymmetry - by a real K\"ahler potential, holomorphic superpotential, holomorphic gauge kinetic functions and Fayet-Iliopoulos D-terms.

\subsection{R-symmetry and Peccei-Quinn symmetry}
The symmetries that underlie the non-renormalisation theorem in the presence of background fluxes and the $\bDt$-brane are an R-symmetry, a PQ-symmety in $\tau$ and a PQ-symmetry in $T$.  The former two are subgroups of the modular symmetry, $SL(2,\R)$, discussed in Section \ref{S:modinv}, with the R-symmetry transformations corresponding to:
\be
\i b = |\tau|, \quad \i c=1/|\tau| \quad \text{and} \quad d=0
\ee
and the PQ-symmetry transformations corresponding to:
\be
a=d=1 \quad \text{and} \quad c=0 \,.
\ee
   As already mentioned, although the $SL(2,\R)$ symmetry is only a symmetry to leading order in $g_s$ and $\alpha'$, quantum corrections are expected to preserve its discrete subgroup $SL(2,\Z)$, once non-perturbative effects are also included.  Moreover, the continuous R-symmetry and PQ-symmetries are preserved to leading order in $\alpha'$ (being symmetries of the 10D supergravity action) and -- at this leading order in $\alpha'$ -- to all orders in $g_s$ (since supersymmetry fixes the dilaton dependence in the 10D supergravity action).  The PQ symmetries in $\tau$ and $T$ descend from 10D RR gauge invariance.    One subtlety in our arguments is that the flux compactification in the presence of a $\bDt$-brane is only a static solution if the runaway volume modulus is stabilised by non-perturbative effects, and so -- for consistency -- we have to incorporate these non-perturbative effects in the 4D LEEFT. However, as the non-renormalisation theorems only apply to perturbative corrections, we will neglect these effects in what follows.

\subsection{Spontaneously broken $SL(2,\R)$ and spurions}

Clearly, when type IIB supergravity is compactified in a GKP background, the $SL(2,\R)$ symmetry -- and its R-symmetry and Peccei-Quinn subgroups -- are spontaneously broken by the non-trivial vacuum expectation values for the fluxes and dilaton (see \eqref{E:SL2Zxfmn}).  Thus the 4D EFT describing the compactification at low energies will not enjoy the $SL(2,\R)$ invariance.  However, we can replace the symmetry breaking parameters -- which correspond to combinations of flux and dilaton {\it v.e.v's} -- with auxiliary spurion fields, $\mathcal{G}^r$, which transform under $SL(2,\R)$ in such a way as to recover the symmetry.  As we know all the sources of symmetry breaking -- flux and constant dilaton vevs -- we can then write down all possible operators in the EFT by constructing the invariant operators and, finally, restoring the spurion field to the original constant parameters. The superpotential becomes expressed in terms of these spurions $\mathcal{G}_r$ and the light fields.

	Using the modular transformation properties given in Section \ref{S:modinv}, we can work out the R-charges for the various objects.  The Grassmanian superspace coordinate, $\theta$, has R-charge $R_\theta = -\frac12$, from which one can deduce that the superpotential has R-charge $R_W = -1$.   The R-charges for the light fields and spurions are:
	\be
	R_{T,Z,Y^i} = 0\,, \quad R_X = -1 \quad \textrm{and} \quad R_{\mathcal{G}_r}=-1 \,.
	\ee
  Then, indeed, the superpotential \eqref{E:KWfinal} has $R_W = -1$.  Meanwhile, all fields (apart from of course $\tau$ and $T$ themselves) are invariant under the PQ-symmetries $\tau \rightarrow \tau + \i \, const$ and $T \rightarrow T + \i \, const$.
 	
\subsection{Proof of the non-renormalisation theorem}
Let us first consider the action to leading order in $\alpha'$.  The action enjoys the R-symmetry discussed above, with the superpotential carring R-charge -1.  Collecting the fields with non-trivial R-charge -1 together, $\mathcal{G}^s=(\mathcal{G}^r\,, X)$, and those with trivial R-charge as $\varphi^\alpha = (T,Z,Y^i)$, the full superpotential can then be written without loss of generality as:
\be
W(\tau\,, \mathcal{G}^s\,, \varphi^\alpha) = \mathcal{G}^0 w(\varphi^\alpha, \mathcal{G}^s/\mathcal{G}^0)
\ee 
for some field $\mathcal{G}^0 \in \{\mathcal{G}^s\}$.  
Note that $w$ cannot depend separately on $\tau$ (beyond its dependence via $\mathcal{G}^r$) as $\tau$ transforms under the PQ-symmetry, whereas $W$ must be PQ-invariant.  However, we cannot argue that $W$ is independent of $\tau$ - since it depends on $\tau$ via $\mathcal{G}^r$ - and therefore we cannot use this independence to argue that $W$ is protected from perturbative string-loop corrections.  This problem arises because the string coupling constant, $e^{\phi}$, is not the loop counting parameter for the 10D type IIB supergravity action.  To see the non-renormalisation of $W$, we have to reorganise the string-loop expansion\footnote{See \cite{BEQ} for the argument as to why we can meaningfully reorganise this generically divergent -- but asymptotic -- loop-expansion.}, by performing the following rescaling of the 10D fields:
\be
e^\phi \rightarrow \lambda e^\phi\,, \quad G_3 \rightarrow \lambda^{-1} G_3\,, \quad \Theta \rightarrow \lambda^{-1/2} \Theta \,. \label{E:rescaling}
\ee
With this rescaling the action (in the string frame) acquires an overall factor $\calS \rightarrow \lambda^{-2} \calS$ so that $\lambda$ can be used a loop-counting parameter for type IIB supergravity.  That is, we may perform the rescaling \eqref{E:rescaling} and dimensionally reduce, whereby:
\be
{\mathcal G}_r \rightarrow \lambda^{-1} {\mathcal G}_r\,, \quad \psi_X \rightarrow \lambda^{-1} \psi_X \quad \textrm{and} \quad X \rightarrow \lambda^{-1} X \,.
\ee
Then we formally expand the 4D EFT in terms of the loop-counting parameter $\lambda$, taking back $\lambda \rightarrow 1$ at the end.  If the superpotential receives no higher order corrections in the supergravity-loop counting parameter $\lambda$, then nor will it receive corrections in the string-loop counting parameter, $g_s$.

Finally, we can easily show that $W$ receives no corrections to all orders in $\lambda$.  Indeed, the arguments of the function $w(\varphi^\alpha, \mathcal{G}^s/\mathcal{G}^0)$ are clearly independent of $\lambda$ and so $W$ receives no corrections in the expansion in $\lambda$.  Therefore, the superpotential, to leading order in the $\alpha'$ expansion, receives no corrections to all orders in the string-loop expansion.  Moreover, the PQ-symmetry $T \rightarrow T + \i \, const$ together with the holomorphy of $W$ ensures that $W$ receives no corrections to all orders in the $\alpha'$ expansion, just as in the original non-renormalisation theorem.

Similar arguments can be applied to the gauge kinetic functions.  In the end, the 4D non-linear $\mathcal{N}=1$ EFT describing string flux compactifications with an $\bDt$-brane can receive perturbative $g_s$ and $\alpha'$ corrections only via the K\"ahler potential and up to one loop in the gauge kinetic functions, whereas the superpotential can only receive non-perturbative corrections.  As it is the superpotential that generally determines the vacuum structure of the theory, this result puts such vacua on much firmer footing.

\section{Discussion}
We have derived the 4D LEEFT for CY orientifold compactifications of type IIB supergravity with supersymmetric background fluxes in the presence of a probe $\bDt$-brane  and non-perturbative effects, including all the bulk moduli.  As the CY orientifold flux geometry and the $\bDt$-brane together spontaneously break supersymmetry (each preserving a different ${\mathcal N}=1$ supersymmetry), at energy scales above the gravitino mass but below the warped string scale associated with the $\bDt$-brane tension, the LEEFT can be written in terms of a non-linearly realised ${\mathcal N = 1}$ supergravity theory. 

Our computations proceeded by writing down a 4D non-linear supergravity Lagrangian -- a real $K$, holomorphic $W$ and $H$ -- whose kinetic, mass and uplift terms match those obtained from dimensionally reduction of the 10D theory.  The non-linear supersymmetry inferred from the symmetry of the antibrane-flux setup can then be used to obtain the remaining couplings in the LEEFT at leading order.  The 4D action is invariant under the 4D modular transformations -- provided we allow background fields to transform -- as it must be.  It also recovers the expected results both in the vanishing antibrane limit and the vanishing flux, decompactification limit.

The non-linear local supersymmetry can be described with the use of constrained superfields appropriately coupled to gravity and matter.  The non-linear supergravity action can be written in several different ways, depending on the kind of constrained superfield in which we choose to place the goldstino fermion of the spontaneously broken local supersymmetry.  A nilpotent chiral superfield, $S$, contains the goldstino and auxiliary F-term field as independent component fields.  The 4D LEEFT for KKLT can then be written as:
\bea
 \kappa_4^{2}\,K &=& - \ln(\tau + \bar \tau)-   \ln f(Z, \bar Z) \nn \\
&& -3  \ln\left(T+\bar T - \kappa_4^2 \frac{e^{-4A_0}(T+\bar T) }{3(\tau+ \bar \tau)f(Z,\bar Z)} S \bS  -  \kappa_4^2 \frac{e^{-4A_0}(T+\bar T)}{3f(Z,\bar Z)} \delta_{i\bar{i}} Y^i \bar Y^{\bar i}\right)  
\eea
and
\be W = \int G_3 \wedge \Omega + M^2 S + h_{ij}(\tau) Y^i Y^j 
\ee
  The uplift of the scalar potential due to the positive energy density contribution from the $\bDt$-brane is here described using an F-term scalar potential associated with the superpotential term $W = M^2 S$.  

We can also parameterise the brane contributions in term of the constrained chiral superfield, $X$, which satisfies both the nilpotency condition and an additional derivative constraint.  The latter is chosen precisely to fix the F-term auxiliary field, in such a way that the F-term scalar potential matches the uplift term from the $\bDt$-brane.  Equivalently, the brane contributions can be written in terms of a constrained vector superfield, $V$, with the uplift potential corresponding to a D-term potential.  This is even though the constraints on $V$ fix all the component fields except for the goldstino, and in particular remove the gauge boson and gauge symmetry.  In the case of non-linearly realised supersymmetry, there is no physical distinction between F-term and D-term uplifting.

Having a 4D LEEFT description of the antibrane-flux setup including all bulk moduli sheds light on several aspects of the KKLT scenario for stringy dS vacua.  Once again, we have evidence that the antibrane spontaneously breaks supersymmetry and that the subsequent non-linearly realised supersymmetry gives us control in the setup.  We can now also avoid the ad-hoc three-step picture of uplifting, where the $\bDt$-brane is added into the picture at the very end, after moduli stabilisation in an adS vacuum by fluxes and non-perturbative effects.  Instead, we can follow the physical hierarchy of scales:
\be
M^{w}_{s} \sim M_{\bcancel{susy}}  \gg  M^{w}_{kk} \sim  \Lambda_{np} \sim \Lambda  \gg M_{\tau,Z}  \gg  M_{T}  \sim M_{3/2} \sim M_{goldstino} 
\ee
to order the dynamics.  The mass of the gravitino is $m_{3/2} = \kappa_4^2 e^{\kappa_4^2 K/2}|\langle W_{np}\rangle|$, with further effective contributions to the super-Higgs mechanism from the Hubble scale associated with the dS vacuum $H^2 = \langle \V \rangle/\Mpl^2$ \cite{FerraraVP,superHiggs}.  

Finally, we have argued that any metastable dS vacuum thus obtained would be robust against quantum corrections, due to the non-renormalisation theorems that we show to hold for the non-linear supergravity theories descending from string compactifications.  These arise thanks to remnants of S-duality and RR-gauge invariance in type IIB supergravity with D3/$\bDt$-branes, and they protect the 4D superpotential from perturbative corrections (as we have seen, non-perturbative corrections can and do come into play).  Indeed, although background fluxes spontaneously break modular invariance, we can follow this breaking by associating these backgrounds with spurion fields.  In this way, we used a combination of an R-symmetry and PQ-symmetries -- together with holomorphicity of the superpotential in the non-linear supergravity action -- to show that the 4D superpotential receives no corrections to all finite orders in the $g_s$ and $\alpha'$ expansions.  The K\"ahler potental will receive perturbative corrections in both expansions, and the superpotential will receive higher order non-perturbative corrections, but the vacuum structure is generally determined by the leading contributions written down here.

There are a number of important directions in which to extend our effective description of $\bDt$-branes in type IIB CY orientifold flux compactifications.  For simplicity, we placed the $\bDt$-brane on top of an O3$^-$-plane, thus projecting out the worldvolume scalars and gauge bosons and leaving only the worldvolume fermions on the brane.  Incidentally, this circumvents any worry about tachyonic instabilities which have been argued to appear when going beyond the probe brane approximation \cite{Bena:2016fqp}.  However, having placed the $\bDt$-brane on an O3$^-$-plane, there does not seem to be a supersymmetric state to which the setup could non-perturbatively decay -- as the antibrane is stuck on the O-plane, it cannot follow the usual KPV \cite{Kachru:2002gs} brane polarisation and flux annihilation (in fact the details of this process is in any case unknown for the case of a single $\bDt$-brane, see however \cite{Polchinski:2015bea}).  Even the decompactification limit leaves the non-supersymmetric $\bDt$/O3$^-$-system, which was argued to be perturbatively stable in \cite{Uranga:1999ib} (perhaps surprisingly given the fact that $\bDt$'s and O3$^-$'s both have negative RR-charge, see also e.g. Section 6.5.3 of \cite{Ibanez:2012zz}).
Placing the $\bDt$-brane away from any O-plane will restore the worldvolume bosonic degrees of freedom, allowing once again non-perturbative decay via antibrane/flux annihilation. Worldvolume bosons should then lie in their own constrained supermultiplets, as proposed by \cite{VW}.  It would be very interesting to extend our analysis to this case, which has been begun in \cite{Aalsma:2017ulu}, as well as the original KKLT scenario where supersymmetry is broken by both the fluxes (with non-trivial $W_0 = \langle \int G_3 \wedge \Omega \rangle$) and $\bDt$-brane.  Work towards understanding the soft-supersymmetry breaking terms for visible sectors with $\bDt$-brane spontaneous supersymmetry breaking was begun in \cite{Aparicio:2015psl}.

\section*{Acknowledgments}
It is our pleasure to thank Fernando Quevedo for inspiring this work and collaboration.  We also acknowledge Nana Cabo Bizet, Mariana Gra\~na,  Daniel Junghans, Renata Kallosh, Fernando Quevedo, Savdeep Sethi, Gianmassimo Tasinato, Thomas Van Riet, Bert Vercnocke and Tim Wrase for helpful discussions.  The work of SLP was supported by a Marie Curie Intra European Fellowship within the 7th European Community Framework Programme.  MPGM is supported by project MECESUP 1655 and by the project Fondecyt 1161192.  SLP and IZ would like to thank DESY, Hamburg, the Michigan Center for Theoretical Physics, Ann Arbor and The Lorentz Center, Leiden for hospitality during the latter stages of this work.

\appendix

\section{Notation and Conventions}

We use a mostly minus signature.  Our conventions for the 10D gamma matrices are:
 \be
	\Gamma^{\mu} = \gamma^{\mu} \otimes \Id_8 \quad (\mu = 0,1,2,3) \quad \textrm{and} \quad \Gamma^{m} = \gamma_{5} \otimes \gamma^{m} \quad (m=4,5,6,7,8,9)
	\ee 
from which it follows that:
 \be
 \Gamma_{11}= \gamma_5 \otimes \gamma_7 \,.
 \ee
For the 4D gamma matrices:
\be
	\label{eq:gamma4d}
	 \gamma^\mu =\begin{pmatrix}
	0 & \sigma^\mu \\
	\bar\sigma^\mu & 0 \\
	\end{pmatrix} \,,
	\ee
	where $\sigma^\mu = (-\Id, \vec\sigma)$ and $\bar\sigma=(\Id, \vec{\sigma})$  where $\Id$ stand for the identity matrix  and $\vec \sigma$  denote Pauli matrices. 
	In particular,
	\be
	\gamma^0 = 	\begin{pmatrix}
	0 & -\Id \\
	\Id & 0 \\
	\end{pmatrix}\,, \hspace{1cm}
		\gamma^5 = i \gamma^0 \gamma^1\gamma^2 \gamma^3 =\begin{pmatrix}
	\Id & 0 \\
	0 & -\Id \\
	\end{pmatrix} \,.
	\ee
	The charge conjugation matrices satisfy:
	\be
	C_{10} = C_4 \otimes C_6 \,, \quad \textrm{with} \quad C_{10}^T = - C_{10}\,, \quad C_4^T = - C_4 \quad \textrm{and} \quad C_6^T = C_6 \,,
	\ee
	and also:
	\be
	\Gamma_M^T = C_{10} \Gamma_M C_{10}^{-1} \,.
	\ee
	A convenient choice for the charge conjugation matrices is:
	\bea
	C_4 = i\gamma^3 \gamma^1=   \begin{pmatrix}
	0& i & 0 & 0\\
	-i & 0 & 0 & 0\\
	0& 0 & 0 &  i\\
	0& 0 &- i& 0
	\end{pmatrix} \quad \textrm{and} \quad
	C_{(6)} = \sigma_1 \otimes \sigma_1 \otimes \sigma_1 =  \begin{pmatrix}
	0& 0 & 0 & 1\\
	0 & 0 & . & 0\\
	0&  . & 0 &  0\\
	1& 0 & 0 & 0
	\end{pmatrix} \,.
	\eea
We use the following definitions for Dirac conjugation, charge conjugation and complex conjugation for a general Dirac spinor:
\be
\bar\Psi \equiv \Psi^\dagger \Gamma^0 \quad \textrm{and} \quad \Psi^C \equiv C_{10} \bar\Psi^T \,.
\ee
4D complex Weyl spinors are:
\be
\label{eq:pnchirality}
\psi_{+} =\begin{pmatrix} 
\psi_\alpha \\
0
\end{pmatrix} 
\hspace{1cm} 
  \psi_{-}= \begin{pmatrix} 
0 \\
\bar\psi^{\dot{\alpha}}
\end{pmatrix} \,,
\ee
from which we can compose a 4D Majorana spinor:
\be
\psi =\begin{pmatrix} 
\psi_\alpha \\
\bar\psi^{\dot{\alpha}}
\end{pmatrix} \,.
\ee

\section{Modular transformation of the worldvolume fermion}

The $\bDt$-brane action \eqref{E:S3} contains the following combination of fields:
\be
\calL^{\bDt}_{2-f} \supset e^{\phi/2}\Im G^{ISD}_{mnp}\,\bar \Theta \Gamma^{mnp} \Theta \,,
\ee
which must therefore be modular invariant.  Converting to complex coordinates:
\be
e^{\phi/2}\Im G^{ISD}_{mnp}\,\bar \Theta \Gamma^{mnp} \Theta = \frac{1}{2\i}\Big( e^{\phi/2} G_{ijk}^{ISD} \bar \Theta \Gamma^{ijk} \Theta - e^{\phi/2} {\bar G}_{\bar i \bar j \bar k}^{ISD} \bar \Theta \Gamma^{\bar i \bar j \bar k} \Theta + \dots \Big) \,,  \label{E:10DGTheta}
\ee
where we have omitted the $ij\bar k$, $i\bar j \bar k$ components for brevity. At this point it is not easy to see how the above term from the brane action can be modular invariant, but progress can be made by dimensionally reducing the 10D MW spinor as:
\be
\Theta = \lambda_-\otimes \zeta_- + \psi_-^i \otimes \zeta^i - C_4 \gamma^0 \lambda_-^* \otimes C_6 \zeta_-^* - C_4 \gamma^0 (\psi_-^i)^* \otimes C_6 (\zeta^i)^*
\ee
where $\zeta_-$ is the 6D nowhere vanishing covariantly constant (w.r.t. the unwarped metric) spinor of negative chirality, which is annihilated by $\gamma^{\bar i} \zeta_-$.  Also, $\zeta_i = \frac{1}{6||\Omega||}\Omega_{ijk} \gamma^{jk} \zeta_-$ has negative chirality.  Dimensional reduction of \eqref{E:10DGTheta} then yields:
\bea
&&\frac{1}{2\i}\Big( e^{\phi/2} G_{ijk}^{ISD} \bar \Theta \Gamma^{ijk} \Theta - e^{\phi/2} {\bar G}_{\bar i \bar j \bar k}^{ISD} \bar \Theta \Gamma^{\bar i \bar j \bar k} \Theta \Big)  \\
&&\quad =  \frac{1}{2\i}\Big(-e^{\phi/2}G_{ijk}^{ISD} \lambda_-^T C_4 \lambda_- \otimes \zeta_-^T C_6 \gamma^{ijk} \zeta_- + e^{\phi/2}{\bar G}_{\bar i \bar j \bar k}^{ISD} {\lambda}_{-}^{\dagger} C_4 \lambda_{-}^{* }\otimes \zeta_-^\dagger \gamma^{\bar i \bar j \bar k} C_6 \zeta_-^* + \dots \Big) \nn
\eea
where several terms vanish due to $\gamma^{\bar i} \zeta_- = 0$, but we have omitted several non-vanishing terms for brevity.  Modular invariance is then satisfied provided that:
\be
\lambda_- \rightarrow e^{-\i \gamma_5 \delta} \lambda_-, \quad \psi^i_- \rightarrow e^{-\i \gamma_5 \delta} \psi^i_-, \quad \lambda_-^* \rightarrow e^{\i \gamma_5 \delta} \lambda_-^*, \quad \textrm{and} \quad (\psi^i_-)^* \rightarrow e^{\i \gamma_5 \delta} (\psi^i_-)^* \,.
\ee
Finally, we see that the modular transformation of $\Theta$ must be:
\be
\Theta \rightarrow e^{-\i  (\gamma_{5} \otimes \Id) \delta} \Theta \,,
\ee
 where $\gamma_5 \otimes \Id$ appears to ensure that the Marjorana condition on $\Theta$ continues to be satisfied after the modular transformation. Notice that if we plugged this transformation into \eqref{E:10DGTheta}, we would not immediately obtain modular invariance due to the second term.  To see the modular invariance, we need to exploit the brane origin of the worldvolume fermion, and decompose it according to its longitudinal and normal directions, where the normal space has structure group $SU(3)$ and covariantly constant spinor $\zeta_-$.

\bibliography{refs}

\bibliographystyle{utphys}

\end{document}